# Review of magnetic properties and magnetocaloric effect in the intermetallic compounds of rare earth with low boiling point metal(s)*

Li Ling-Wei (李领伟)[†]

*Key Laboratory of Electromagnetic Processing of Materials (Ministry of Education), Northeastern University, Shenyang 110819, China*

**Abstract:** The magnetocaloric effect (MCE) in many rare earth (*RE*) based intermetallic compounds has been extensively investigated during last two decades, not only due to their potential applications for magnetic refrigeration but also for better understanding of the fundamental problems of the materials. This paper reviews our recent progress of magnetic properties and MCE in some binary or ternary intermetallic compounds of *RE* with low boiling point metal(s) (Zn, Mg, and Cd). Some of them are exhibiting promising MCE properties which make them also attractive for low temperature magnetic refrigeration. Characteristics of the magnetic transition, origin of large MCE as well as the potential application of these compounds were thoroughly discussed. Additionally, a brief review of the magnetic and magnetocaloric properties in the quaternary rare earth nickel boroncarbides *RE*Ni$_2$B$_2$C superconductors is also presented.

**Keywords:** Magnetocaloric effect; Rare earth based intermetallic compounds; Low boiling point metal(s); *RE*Ni$_2$B$_2$C superconductors; Magnetic phase transition

* Project supported by the National Natural Science Foundation of China (Nos. 11374081 and 11004044), the Fundamental Research Funds for the Central Universities (Grant Nos. L1509006 and N140901001), the Japan Society for the Promotion of Science Postdoctoral Fellowships for Foreign Researchers (No. P10060), and the Alexander von Humboldt (AvH) Foundation (Research stipend to L. Li). I sincerely thank Prof. Rainer Pöttgen from Universität Münster, Germany and Prof. Katsuhiko Nishimura from University of Toyama, Japan for the fruitful discussions and helpful suggestions.

[†] Author to whom correspondence should be addressed. Electronic mail: lingwei@epm.neu.edu.cn (Lingwei Li), wei0396@hotmail.com (Lingwei Li); Tel. +86-24-83684653.

## 1. Introduction of magnetocaloric effect and materials

The magnetocaloric effect (MCE) is a magneto-thermodynamic phenomenon which was first discovered in pure Fe by Emil Warburg in 1881,[1] and its origin was independently explained by Debye and Giauque in the 1920s.[2, 3] All magnetic materials exhibit MCE, although the intensity of this effect depends on the properties of each material. Generally, for a simple ferromagnetic material near its Curie temperature, when a magnetic field is applied, the spins tend to align parallel to the magnetic field which lowers the magnetic entropy. To compensate for the loss in the magnetic entropy in an adiabatic (isentropic) process the temperature of the material increases. When the magnetic field is removed, the spins tend to become random which increases the magnetic entropy and the material cools. Meanwhile, Debye and Giauque have also proposed the adiabatic demagnetization in order to achieve temperature lower than that of the liquid helium. This possibility was later experimentally demonstrated by the chemist Nobel Laureate William F. Giauque in 1933, he was able to reach a temperature of 0.25 K based on the adiabatic demagnetization of 61g paramagnetic salts $Gd_2(SO_4)_3 \cdot 8H_2O$.[4] Besides the success in the ultralow temperature, the magnetic refrigeration technology has also been applied at the higher temperature ranges, such as ~2 to 20 K, the range of liquid helium and hydrogen, as well as ~20 to 80 K, the range of liquid hydrogen and nitrogen.[5-8]

In 1997, two key developments enhanced the feasibility for producing a magnetic refrigerator for commercial or industrial use. The first is that Zimm et al [9] realized the first near room temperature magnetic refrigeration prototype at the Ames Laboratory. Second is that Pecharsky and Gschneider [10] observed a giant MCE in the $Gd_5(Si_2Ge_2)$ compound. Besides the discovery of the giant MCE in $Gd_5(Si_2Ge_2)$ and related compounds,[10, 11] a number of materials with giant/large MCE have been realized in last three decades, for example $RECo_2$ ($RE$ = Er, Ho, and Dy) alloys,[12, 13] MnAs based compounds,[14-17] manganites (RE,M)MnO$_3$, (RE = lanthanide, M = Ca, Sr, and Ba),[18,19] Ni-Mn-X (X = Ga, In, and Sn) based Heusler alloys,[20-24] La(Fe,Si)$_{13}$ and related compounds,[25-28] MnTX (T = Co, Ni, and Fe; X = Si, Ge) based compounds [29-32] as well as some rare earth (RE) based intermetallic compounds.[33-36] The structure, physical properties as well as the MCE and its application in most of the above mentioned materials have been summarized in a number of reviews articles [11-14, 17-19, 22, 23, 26, 27, 32-35, 37-40] which will not be repeated here. In this paper, we give a brief review of our recent progress in exploring magnetocaloric materials, mainly related to several typical systems of RE-based intermetallic compounds.

## 2. Characterization of magnetocaloric materials

### 2.1. Evaluation of magnetic entropy change ($\Delta S_M$) and adiabatic temperature change ($\Delta T_{ad}$)

The MCE is a intrinsic phenomenon for magnetic materials which manifests as the isothermal magnetic entropy change $\Delta S_M$ or/and the adiabatic temperature change $\Delta T_{ad}$ when it is exposed to a varying magnetic field, the definition of $\Delta S_M$ and $\Delta T_{ad}$ can be found in Fig. 1. The MCE can be measured directly or it can be calculated indirectly from the experimentally measured heat capacity and/or magnetization. In the present review, the reported values of $\Delta S_M$ and $\Delta T_{ad}$ were estimated from two different methods using the indirect technology: (1) from field and temperature dependence of heat capacity $C(H, T)$, (2) from field and temperature dependence of magnetization $M(H, T)$ and zero field heat capacity $C(T, 0T)$ which were described as below:

The temperature dependence of total entropy $S$ under different fields can be calculated from heat capacity by numerical integration:

$$S(H,T) = \int_0^T \left(\frac{C(H,T)}{T}\right)_H dT \tag{1}$$

where $H$ is the magnetic field at which the heat capacity was measured. Then the $\Delta S_M$ as well as the adiabatic temperature change $\Delta T_{ad}$ can be calculated from the heat capacity measurements using the following relations:

$$\Delta S_M(T, \Delta H) = \int_0^T \frac{C(T,H_1) - C(T,H_0)}{T} dT = [S(T)_{H_1} - S(T)_{H_0}]_T, \tag{2}$$

$$\Delta T_{ad}(\Delta H, T) = [T(S)_{H_1} - T(S)_{H_0}]_S. \tag{3}$$

According to thermodynamical theory, the isothermal magnetic entropy changes associated with a magnetic field variation is given by

$$\Delta S_M(T, \Delta H) = S_M(T, H_1) - S_M(T, H_0) = \int_0^H \left(\frac{\partial S(H,T)}{\partial H}\right)_T dH \tag{4}$$

From the Maxwell's thermodynamic relation:

$$\left(\frac{\partial S(H,T)}{\partial H}\right)_T = \left(\frac{\partial M(H,T)}{\partial T}\right)_H, \tag{5}$$

one can obtain the following expression:

$$\Delta S_M(T, \Delta H) = \int_0^H \left(\frac{\partial M(H,T)}{\partial T}\right)_H dH \tag{6}$$

where $S$, $M$, $H$, and $T$ are the magnetic entropy, magnetization of the material, applied magnetic field, and the temperature of the system, respectively. From the magnetization measurements made at discrete field and temperature intervals, $\Delta S_M$ can be approximately calculated by the following expression:

$$\Delta S_M(T, \Delta H) \approx \frac{1}{\delta T}\left[\int_0^H M(T+\delta T, H)dH - \int_0^H M(T, H)dH\right]. \qquad (7)$$

From zero field $C(T)$, we can calculate $S(T, 0\text{ T})$ by using equation 1. Using the magnetic entropy change $\Delta S_M$ calculated from $M(H, T)$, the $S(T, H)$ can be calculated from the following expression:

$$S(T, H_1) = \Delta S_M(T, \Delta H) + \int_0^T \left(\frac{C(T, H_0)}{T}\right)_H dT. \qquad (8)$$

Then the adiabatic temperature change $\Delta T_{ad}$ can be calculated using Eq. 3. Using the present method, the accuracy of the $\Delta S_M$ calculated from the magnetization data for the materials studied here is better than 10%.

## 2.2. Evaluation of the relative cooling power (*RCP*) and refrigerant capacity (*RC*)

Another important quality factor(s) for magnetocaloric materials is the relative cooling power (*RCP*) or/and refrigerant capacity (*RC*) which is a measurement of the amount of heat transfer between the cold and hot reservoirs in an ideal refrigeration cycle. The *RCP* is defined as the product of the maximum magnetic entropy change $-\Delta S_M^{max}$ and full width at half maximum, $\delta T_{FWHM}$, in $\Delta S_M(T)$ curve,

$$i.e.,\quad RCP = -\Delta S_M^{max.} \times \delta T^{FWHM}. \qquad (9)$$

Whereas, the *RC* is calculated by numerically integrating the area under the $-\Delta S_M - T$ curve at half maximum of the peak taken as the integration limits, where $T_1$ and $T_2$ are the temperatures of the cold end and the hot end of an ideal thermodynamic cycle, respectively,

$$i.e.,\quad RC = \int_{T_1}^{T_2} |\Delta S_M| dT. \qquad (10)$$

An example for the evaluation of the *RCP* and *RC* is displayed in Fig. 2. It is easy to find that the value of *RC* is around 25 % lower than that of the *RCP* for the $\Delta S_M(T)$ that with a single broad triangle peak around $T_C$, which is a typical character for the second order MCE materials.

**2.3. Determination of the order of magnetic phase transition**

It is well known that the magnitude, temperature and magnetic field change dependence of MCE have strong correlation with the nature of the corresponding magnetic phase transition; therefore, it is important to determine the order of magnetic transition of the magnetocaloric materials. Based on the Inoue-Shimizu model,[41, 42] which involves a Landau expansion of the magnetic free energy up to sixth power of the total magnetization *M*, can be used to determine the magnetic transition type,[41,42]

$$F(M,T) = \frac{c_1(T)}{2}M^2 + \frac{c_2(T)}{4}M^4 + \frac{c_3(T)}{6}M^6 + \cdots - BM . \qquad (11)$$

After mimimization of the above equation with respect to the magnetization M, the magnetization near the Curie temperature $T_C$:

$$c_1(T)M + c_2(T)M^3 + c_3(T)M^5 = B . \qquad (12)$$

The parameters $c_1(T)$, $c_2(T)$ and $c_3(T)$ represent the Laudau coefficient, and it has been reported that the order of a magnetic transition is related to the sign of the $c_2(T)$. A transition is expected to be the first order when $c_2(\sim T_C)$ is negative, whereas it will be the second order for a positive $c_2(\sim T_C)$. The sign of $c_2$ can be determined by Arrott plot (*H*/*M* versus $M^2$ curves)[43], *i. e.* the magnetic transition is of second order if all the *H*/*M* versus $M^2$ curves have positive slope. On the other hand, if the *H*/*M* versus $M^2$ curves show negative slope, the magnetic transition is of the first order. In the present review, the order of magnetic phase transition of all the materials was determined by this method.

## 3. MCE in rare earth based intermetallic compounds that contains low boiling point metal(s)

The rare earth (*RE*) – transition metal (*T*) based systems have intensively been investigated in recent years with respect to their interesting chemical and physical properties as well as practical applications in a wide spectrum of industries. Depending on the constituent element and composition, various properties like magnetic ordering, superconductivity, heavy-fermion behaviour, valence fluctuation, large magneto-resistance (MR) effect and large/giant MCE, *etc*. have been observed.[44-48] Compared to the other transition metals, the research related to Zn, Mg, and Cd-based intermetallic compounds are not too much. This might be a consequence of the difficulty in materials synthesis due to the low boiling point (high vapour pressure) of Zn, Mg, and Cd. Pöttgen et al.[49, 50] have proposed a simple method to prepare such materials. First, stoichiometric amounts (0.5%~1% of the low boiling point metals were added) of the consistent elementals with purity better than 99.9% were weighted. Second, the weighted metals are arc-welded[49] in a Ta or Nb tube under an argon pressure of ca. 75-85 kPa. The argon was purified over titanium sponge (900 K), silica gel, and molecular sieves. Third, the crucible was placed in a water-cooled sample chamber of an induction furnace (Hüttinger Elektronik, Freiburg, Germany, Typ TIG 1.5/300)[50] and heated up to a certain high temperature for several minutes, following by several hours annealing at a selected temperature. Then the obtained samples were grinded and cold-pressed into pellet, and finally annealed at a certain temperature for several days in evacuated quartz tubes. The structural properties and phase purity were subsequently characterized by X-ray powder diffraction (XRD) and Energy Dispersive X-Ray Spectroscopy (EDX) analyses. In the following section, our recent progress related to the magnetism and MCE in the binary or ternary intermetallic compounds of *RE* with low boiling point metal(s) (Zn, Mg, and Cd) is reviewed, respectively. For a quick comparison, the magnetic transition temperature together with MCE parameters for the compounds in the present review as well as some other rare earth based intermetallic compounds with significant MCE properties are summarized in Table I.

### 3.1. Zn-based intermetallic compounds

The equiatomic binary compounds REZn (RE is a rare earth) with the cubic CsCl-type structure (space group $Pm3m$) have attracted some attention due to their simple crystal structure as well as their interesting physical and chemical properties.[51-54] The magnetic studies revealed that the heavy REZn compounds order ferromagnetically with the ordering temperature ranging from 267 K for GdZn to 8.5 K for TmZn, whereas the light ones order antiferromagnetically.[51-54] Sousa et al.[55, 56] have theoretically investigated the MCE in REZn (RE = Tb, Ho, and Er), and some anomalies in $\Delta S_M(T)$ curves are found due to spontaneous and/or spin reorientation transitions in these compounds. Very recently, we have experimentally investigated the magnetic phase transition and magnetocaloric properties of REZn (RE = Tm and Ho) compounds.[57, 58]

Figure 3 shows the temperature dependence of the zero field cooling (ZFC) and field cooling (FC) magnetization (M) for TmZn under various magnetic fields (H) from 0.1 to 1 T. Both ZFC and FC M(T) curves show a typical PM to FM transition. With increasing magnetic field, $T_C$ increases gradually from 8.4 K for H = 0.1 T to 10.3 K for H = 1 T. To evaluate the MCE in TmZn, a set of M(H) curves with increasing and decreasing magnetic field up to 7 T together with the temperature dependence of the heat capacity C for TmZn for the magnetic field changes of 0-2 and 0-5 T for TmZn were measured. No obvious hysteresis can be observed for the whole temperature range. Several isotherms with increasing field are presented in Fig. 4 and the corresponding Arrott plot (H/M versus $M^2$) curves are shown in Fig. 5. For the low temperature ones, the M(H) curves show the typical behavior of a soft ferromagnet. With increasing temperature higher than 8 K, a field induced metamagnetic transition can be observed in the $M(H)$ curves with a wide temperature range, and resulting in a clear S-shape in the Arrott plots which is a typical characterization for the first-order magnetic phase transition.[43] Additionally, the observed magnetic properties in TmZn are quite similar to one of the most important class of MCE materials, $RE$Co$_2$ ($RE$ = Dy, Ho, and Er), which were highlighted due to the occurrence of IEM as well as its close relationship with giant MCE.[12, 13]

The temperature dependence of -$\Delta S_M$ and $\Delta T_{ad}$ for TmZn are shown in Fig. 6 and Fig. 7 which was calculated from the $M$ ($H$, $T$) data and from $C$ ($T$, $H$), respectively. Both $\Delta S_M$ and $\Delta T_{ad}$ obtained by using the two methods as described in Section 2 are in good agreement with each other. The values of -$\Delta S_M^{max}$ are 26.9 and 29.7 J/kg K for the magnetic field changes of 0-5 and 0-7 T, respectively, and the corresponding values of are 8.6 and 11.2 K, indicating that the binary equiatomic compound TmZn belongs to a class of giant magnetocaloric materials. Moreover, for the relatively small field change of 0-1 and 0-2 T, the values of -$\Delta S_M^{max}$ reach 11.8 and 19.6 J/kg K which is beneficial for application. The observed giant MCE in TmZn is related to the first order, field-induced metamagnetic phase transition. I. e., the field sensitive magnetic transition, sharp changes of magnetization around $T_C$ and large values of magnetization of TmZn are considered to be the origin of the giant MCE. For the magnetic field changes of 0-2, 0-5 and 0-7 T, the values of $RCP$ and $RC$ are determined to be 76, 269 and 422 J/kg, and to be 59, 214 and 335 J/kg, respectively. The corresponding values of $\Delta T_{ad}^{max}$ are estimated to be 3.3, 8.6 and 11.2 K.[57] These MCE parameters for TmZn (see Table I), especially for the magnetic field change of 0-2 T, are comparable or obviously larger than most of the potential promising magnetic refrigerant materials in the similar temperature region, the applicability of TmZn attractive for low temperature active magnetic refrigeration.[57]

Whereas, two successive magnetic transitions are observed around $T_C$ ~ 72 K and $T_{SR}$ ~ 26

K for HoZn which is corresponding to paramagnetic to ferromagnetic and spin reorientation transitions, respectively.[58] Magnetization and modified Arrott plots indicate that HoZn undergoes a second-order magnetic phase transition around $T_C$. The critical behaviour around $T_C$ has been performed by modified Arrott plots, scaling analysis and scaling plots and field change dependence on the magnetic entropy change method. The values of critical exponents β, γ and δ for HoZn are determined to be 0.47±0.02, 1.09±0.03 and 3.32±0.04, respectively which have some small deviations from the mean-field theory, indicating a short range or a local magnetic interaction which is properly related to the coexistence of FM and SR transitions at low temperature.[58]

The magnetic entropy change $\Delta S_M$ for HoZn was calculated from $M$ ($H$, $T$) data (as shown in Fig. 8). One large peak together with a shoulder can be observed in the -$\Delta S_M(T)$ curves around the FM and SR transition temperatures, respectively. Two peaks (shoulder) overlap with each other, resulting in a wide temperature range with a large $RCP$ which is beneficial for active applications. The maximum values of the magnetic entropy change (-$\Delta S_M^{max}$) of HoZn for the magnetic field changes of 0-2, 0-5 and 0-7 T are 6.5, 12.1 and 15.2 J/kg K, respectively. The corresponding values of $RCP$ are evaluated to be as large as 255, 792 and 1124 J/kg.[58] These values are obviously larger than those of recently reported for potential magnetic refrigerant materials with similar working temperature under the same field change (see Table I). The large values of -$\Delta S_M^{max}$ and $RC$ indicate that HoZn could be a promising candidate for active magnetic refrigeration in the temperature range of 30-90 K.[58]

Hermes et al.[59] have studied the magnetic properties and MCE in the EuRh$_{1.2}$Zn$_{0.8}$ which crystallizes with the MgCu$_2$ structure and orders ferromagnetically at $T_C$ ~ 95 K. For the magnetic changes of 0-2 and 0-5 T, the values of -$\Delta S_M^{max}$, $\Delta T_{ad}^{max}$ and $RCP$ are evaluated to be 3.1 and 5.7 J/kg K, to be 1.4 and 2.8 K, and to be 233 and 513 J/kg, respectively. A large reversible MCE was reported in EuAuZn around 52 K accompanied by a second order magnetic phase transition from a PM to a FM state.[60] The values of -$\Delta S_M^{max}$ for EuAuZn reach 4.8, 9.1 and 11.3 J/kg K for the magnetic field changes of 0-2, 0-5 and 0-7 T, respectively, and show no thermal and magnetic hysteresis around $T_C$. The corresponding values of $\Delta T_{ad}^{max}$ are evaluated to be 2.1, 3.8 and 4.7 K. Very recently, Zhang et al. [61] have reported the magnetic and magnetocaloric properties in the equiatomic intermetallic compounds of TmZnAl. And they found that the TmZnAl compound undergoes a second order magnetic transition from PM to FM state around 2.8 K. A reversible MCE with considerable magnitude was observed at low temperature. For the magnetic field changes of 0-2, 0-5 and 0-7 T, the maximum values of -$\Delta S_M$ are 4.3, 9.4 and 11.8 J/kg K and the corresponding values $RCP$ ($RC$) are determined to be 53 (41), 189 (149) and 289 (228) J/kg, respectively, for TmZnAl.

### 3.2. Mg-based intermetallic compounds

The ternary *RE*-rich intermetallic compounds of the general composition $RE_4T$Mg ($T$ = Rh, Pd, or Pt), crystallizing in the cubic Gd$_4$RhIn-type structure have attracted some attention due to their interesting physical and chemical properties.[62-65] For this type of crystal structure, there are three crystallographically independent *RE* sites and the rare structural motif of Mg$_4$ tetrahedra. The transition metal atoms are located in $RE_6$ trigonal prisms with strong covalent *RE-T* bonding. These $RE_6T$ basic building units are condensed via common corners and edges to a three-dimensional network in which the cavities are filled by the Mg$_4$ tetrahedra. Recently, we have reported the magnetic properties and MCE of $RE_4$PdMg (*RE* = Eu and Er) and $RE_4$PtMg (*RE* = Ho and Er).[66-68] The magnetic phase transition, the origin of the MCE and its potential application for magnetic refrigeration of these compounds will be shown as below.

Figure 9 (a) shows the temperature dependent ZFC and FC magnetization *M* (left scale) and d$M_{FC}$/d$T$ (right scale) under $H$ = 0.1 T for Eu$_4$PdMg. No difference is observed between the ZFC and FC *M-T* curves around $T_C$ which is usual in magnetic materials with a second order magnetic transition. Only one peak in d*M*/d*T* vs. *T* is observed around 150 K which is corresponding to the PM to FM magnetic transition. Figure 9 (b) shows the FC magnetization *M* (left scale) and d$M_{FC}$/d*T* (right scale) for Eu$_4$PdMg under *H* = 1 T. We can note that *M* increases continuously with decreasing temperature, and d*M*/d*T* shows a table-like behaviour from 20 to 150 K.[66] Fig. 10 shows the *M*(*T*) curves for Eu$_4$PdMg under various magnetic fields up to 7 T. The magnetic transition is very sensitive to the magnetic field, a sharp PM to FM transition can be observed under low magnetic field, and the transition getting broader gradually with increasing magnetic field. The magnetization almost shows linear temperature dependence under higher magnetic fields. This exotic magnetic transition behaviour may be related to its special crystal structure and abundant rare earth site occupation.[66]

A set of magnetic isotherms were measured for Eu$_4$PdMg in the temperature range from 3 to 220 K to evaluate its magnetic entropy change. No obvious hysteresis can be observed over the temperature range. Additionally, the order of magnetic phase transition for Eu$_4$PdMg is confirmed to be second order based on the Banerjee criterion. The resulted -$\Delta S_M$(*T*) curves with the magnetic field changes up to 0-7 T is shown in Fig 11. The -$\Delta S_M$(*T*) curves show a pronounced peak around $T_C$ for a low magnetic field change which is similar to the typical behaviour for magnetocaloric materials with a single magnetic phase transition.[66] A very broad table-like behaviour can be observed for a high magnetic field change which is beneficial for application. These peculiar MCE properties in Eu$_4$PdMg are probably related to the magnetic field sensitive magnetic phase transition and large saturated magnetic moment which is corresponding to its exotic magnetic and/or crystal structure.[66] With the magnetic field changes of 0-2, 0-5 and 0-7 T, the values of -$\Delta S_M^{max}$ are 2.6, 5.5 and 7.2 J/kg K, respectively. The corresponding values *RCP* (*RC*) are determined to be 53 (41), 189 (149) and 289 (228) J/kg. As a matter of fact, the values of -$\Delta S_M^{max}$ are smaller than those of recently reported giant MCE materials at the low temperature region which is related to the additional internal entropy loss due to the larger lattice heat capacity of the materials with the higher $T_C$. The -$\Delta S_M^{max}$ value of Eu$_4$PdMg is comparable with those of some potential magnetic refrigerant materials in the similar temperature region under the same field change (see table I). The *RC* and *RCP* values are obviously larger than those of some potential magnetic refrigerant materials at similar temperature region. The excellent MCE properties indicate that Eu$_4$PdMg is a promising candidate for active magnetic refrigeration in the 20 to 160 K temperature range. The present results may also provide an important clue for searching suitable refrigerant materials in the category of materials that with a field sensitive magnetic phase transition(s).[66]

Very recently, the magnetic properties and the MCE in Ho$_4$PtMg, Er$_4$PdMg, and Er$_4$PtMg compounds have been investigated by magnetization and heat capacity measurements.[67-68] The compounds undergo a second order magnetic phase transition from a PM to FM state at Curie temperatures of $T_C$ ~ 28, 21 and 16 K for Ho$_4$PtMg, Er$_4$PtMg, and Er$_4$PdMg, respectively. The resulted temperature dependence of -$\Delta S_M$ and $\Delta T_{ad}$ with various magnetic field changes up to 0-7 T for Ho$_4$PtMg, Er$_4$PdMg and Er$_4$PtMg are shown in Fig. 12 (a), (b) and (c), and in Fig. 13 (a), (b) and (c), respectively. With the magnetic field changes of 0-1, 0-2, 0-5 and 0-7 T, the values of -$\Delta S_M^{max}$ are evaluated to be 2.8, 6.1, 13.4 and 16.9 J/kg K, to be 2.8, 6.2, 15.5 and 20.6 J/kg K, and to be 3.9, 8.5, 17.9 and 22.5 J/kg K for Ho$_4$PtMg, Er$_4$PdMg and Er$_4$PtMg, respectively, indicating that the presently studied compounds belong to a class of large MCE materials. For the magnetic field changes of 0-2, 0-5 and 0-7 T, the values of *RCP* are determined to be 177, 527 and 762 J/kg; 142, 457 and 742 J/kg K; 152, 483 and 716 J/kg K for Ho$_4$PtMg, Er$_4$PdMg and Er$_4$PtMg, and the corresponding values of $\Delta T_{ad}^{max}$ are evaluated to be 1.8, 4.1 and 5.3 K, to be 1.7, 3.7 and 5.5 K, and to be 2.2, 5.0 and 6.5 K, respectively. The values for the presently studied compounds are comparable or even larger than some of the potential promising magnetic refrigerant materials in the similar temperature region, classifying them as considerable candidates for low temperature magnetic refrigeration.[67, 68]

Linsinger et al.[69] have studied magnetism and MCE in the $Gd_2Ni_xCu_{2-x}Mg$ ($x$ = 1 and 0.5) compounds. Both compounds show a large reversible MCE near their ordering temperatures. The values of -$\Delta S_M^{max}$ reach 9.5 and 11.4 J/kg K for the field change of 0-5 T with no obvious hysteresis loss around 65 K for x = 0.5 and x = 1, respectively. The corresponding values of *RCP* reach 688 and 630 J/kg which is relatively high compared to other MCE materials in that temperature range. These results indicate that $Gd_2Ni_xCu_{2-x}Mg$ could be a promising system for magnetic refrigeration at temperatures below liquid $N_2$. The values of $\Delta T_{ad}^{max}$ for $Gd_2Ni_{0.5}Cu_{1.5}Mg$ reach 1.6 and 3.2 K for the magnetic field changes of 0-2 and 0-5 T, respectively, whereas $\Delta T_{ad}^{max}$ for $Gd_2NiCuMg$ is slightly higher with 1.8 K and 4.3 K. Gorsse et al., have investigated the MCE in $Gd_4Co_2Mg_3$ compound.[70] They found that the compound orders antiferromagnetically below $T_N$ = 75 K and shows a field-induced transition at approximately 0.93 T at 6 K. The values of -$\Delta S_M^{max}$ reach 5.8 and 10.3 J/kg K at 77 K for the magnetic field changes of 0-2 and 0-4.6 T, respectively, which is related to the field-induced magnetic phase transition. The corresponding values of $\Delta T_{ad}^{max}$ are 1.3(1) and 3.4(1) K, respectively.

### 3.3. Cd-based intermetallic compounds

Due to the high toxicity of Cd and its compounds, the Cd-based alloys and compounds have limited technical application, and there is little work on the related topics.[71-73] Hermes et al.[72] have investigated the magnetic and MCE properties in $Er_4NiCd$ compound which crystallizes in the $Gd_4RhIn$ type structure. They found that $Er_4NiCd$ shows Curie-Weiss behavior above 50 K with $T_N$ = 5.9 K. At field strength of 0.4 T a metamagnetic step is visible, together with the positive paramagnetic Curie-temperature (7.5 K) indicative for the rather unstable antiferromagnetic ground state. The values of $\Delta S_M^{max}$ reach -7.3 and -18.3 J/kg K, and the resulted *RCP* are 237 and 595 J/kg, for the magnetic field changes of 0-2 and 0-5 T, respectively, for Er$_4$NiCd. The corresponding values of $\Delta T_{ad}^{max}$ are 3.1 and 7.7 K for the magnetic field changes of 0-2 and 0-5 T, respectively. These results indicate that Er$_4$NiCd could be a promising system for magnetic refrigeration at temperatures below liquid H$_2$. [72]

Recently, Li et al. [73] have investigated the magnetism and MCE in GdCd$_{1-x}$Ru$_x$ ($x$ = 0.1, 0.15, and 0.2) solid solutions. Figure 14 shows the temperature dependence of ZFC and FC magnetization for GdCd$_{1-x}$Ru$_x$ under a low magnetic field of $H$ = 0.1 T. A typical PM to FM transition can be observed around $T_C$ ~ 149, 108 and 73 K for $x$ = 0.1, 0.15 and 0.2, respectively. No obvious thermal hysteresis can be observed between ZFC and FC *M-T* curves which is benefit for application. To evaluate the MCE in GdCd$_{1-x}$Ru$_x$, a set of *M(H)* curves around its own $T_C$ were measured. The resulted temperature dependence of -$\Delta S_M$ for $x$ = 0.1, 0.15 and 0.2 are shown in Fig. 15 (a), (b) and (c), respectively. The -$\Delta S_M$ ($T$) curves show pronounced peaks around its own $T_C$, and the peak is getting sharper and the peak temperature shifts to lower temperature gradually with increasing $x$. Interestingly, a table-like behaviour in -$\Delta S_M$ ($T$) curve for $x$ = 0.1 can be observed which is benefit for application, especially for a single material. For the magnetic field changes of 0-2, 0-5 and 0-7 T, the values of -$\Delta S_M^{max}$ are evaluated to be 1.8, 4.1 and 5.6 J/kg K; to be 2.4, 5.8 and 7.8 J/kg K; and to be 3.8, 8.5 and 11.0 J/kg K for $x$ = 0.1, 0.15 and 0.2, respectively.

The -$\Delta S_M^{max}$ as a function of magnetic field change $\Delta H^{2/3}$ for GdCd$_{1-x}$Ru$_x$ is shown in the inset of Fig. 15 (c). Franco et al. [74, 75] have previously proposed a universal relation in a magnetic system with second-order phase transition between the -$\Delta S_M^{max}$ and $\Delta H$, -$\Delta S_M^{max}$ ≈ A$(\Delta H)^n$, where A is a constant. A linear relation between -$\Delta S_M^{max}$ and $(\Delta H)^{2/3}$ can be observed for all the present studied samples under high magnetic field changes, indicating the critical behaviour for GdCd$_{1-x}$Ru$_x$ is close to the mean field theory. For the magnetic field change of 0-5 T and 0 to 7 T, the values of RCP for GdCd$_{1-x}$Ru$_x$ are determined 636, 597 and 583 J/kg, and to be 889, 852 and 828 J/kg for $x$ = 0.1, 0.15 and 0.2, respectively. Additionally, the GdCd$_{1-x}$Ru$_x$ solid solutions are quite interesting, since by changing the Ru content, it can cover a wide temperature range with excellent MCE properties and then the materials can be combined to design a composite material used as a magnetic refrigerant. Therefore, the considerable $\Delta S_M^{max}$ without any thermal/magnetic hysteresis and large RCP values together with the tunable $T_M$ indicate that the GdCd$_{1-x}$Ru$_x$ solid solutions are attractive for active magnetic-refrigeration.[73]

## 4. Magnetic properties and MCE in $RE$Ni$_2$B$_2$C superconductors

The quaternary rare earth nickel boroncarbides $RE$Ni$_2$B$_2$C ($RE$ = rare earth) which crystallize in the tetragonal LuNi$_2$B$_2$C-type structure have attracted much attention due to the coexisting of superconductivity and magnetic ordering phenomena.[99-103] The compounds are long range magnetic ordered for $RE$ = Gd, Tb, Dy, Ho, Er, and Tm. And the superconductivity is observed for $RE$ = Y, Lu, Dy, Ho, Er and Tm. I. e., superconductivity coexists with magnetic order for $RE$ = Dy, Ho, Er, and Tm, and the ratio of superconducting transition temperature ($T_{SC}$) to antiferromagnetic (AF) ordering temperature $T_N$ ranges from $T_{SC}/T_N$ = 7.0 for TmNi$_2$B$_2$C to 0.60 for DyNi$_2$B$_2$C systems.[101-103] The coexistence and competition of magnetism and superconductivity in these compounds have been systematically investigated experimentally and theoretically.[99-103] In recent years, we have further investigated the MCE in parent and doped $RE$Ni$_2$B$_2$C superconductors.[104-107] The main achievements will be shown in the following sections. The superconducting transition and magnetic ordering temperatures together with MCE parameters are summarized in Table II for a quick comparison of the MCE in the $RE$Ni$_2$B$_2$C superconductors.

## 4.1. Giant reversible MCE in Dy$_{0.9}$Tm$_{0.1}$Ni$_2$B$_2$C superconductor

Figure 16 shows the low temperature magnetic isothermals on increasing (open symbols) and decreasing (filled symbols) field for Dy$_{0.9}$Tm$_{0.1}$Ni$_2$B$_2$C at 2 and 3 K as well for DyNi$_2$B$_2$C at 2 K for a comparison. A large hysteresis for DyNi$_2$B$_2$C was observed, in contrast, the hysteresis of Dy$_{0.9}$Tm$_{0.1}$Ni$_2$B$_2$C was ignorable for $T ≥ 3$ K. The magnetic phase transition $T_M$ and superconducting transition temperature $T_{SC}$ are determined to be 9.2 and 4.5 K, respectively, which are lower than those in DyNi$_2$B$_2$C.[102, 104] A set of selected magnetic isothermals $M(H)$ were measured to evaluate the MCE in Dy$_{0.9}$Tm$_{0.1}$Ni$_2$B$_2$C. The $M(H)$ curves were converted to the Arrot-plot of $H/M$ vs. $M^2$ for Dy$_{0.9}$Tm$_{0.1}$Ni$_2$B$_2$C at some selected temperatures which reveals the occurrence of a first order magnetic phase transition because there is a clear S-shaped behavior below the transition temperature, based on the Banerjee criterion.[43]

The resulted magnetic entropy changes $-\Delta S_M$ which was calculated from $M(H, T)$ curves in Dy$_{0.9}$Tm$_{0.1}$Ni$_2$B$_2$C as a function of temperature are shown in Fig. 17. The temperature and magnetic field change dependences of $-\Delta S_M(T)$ curves provide valuable information on the magnetic ordering in Dy$_{0.9}$Tm$_{0.1}$Ni$_2$B$_2$C. For the magnetic field changes $\Delta H ≤ 2$ T, $-\Delta S_M$ is negative (inverse MCE) below the transition temperature, and changes to positive with increasing temperature, which is corresponding to the magnetic transition from PM to AFM state. For $\Delta H ≥ 3$ T, a positive $-\Delta S_M$ (i.e. conventional MCE) with a broad peak around 13 K is observed. The values of maximum magnetic entropy change ($-\Delta S_M^{max}$) reach 14.7 and 19.1 J/kg K for the magnetic field change of 0-5 and 0-7 T, respectively. The giant MCE was believed to be related to the field induced first-order magnetic transition from AFM to FM.[104] The field dependence of $-\Delta S_M^{max}$ for Dy$_{0.9}$Tm$_{0.1}$Ni$_2$B$_2$C is also shown in the inset of Fig 17. The value of $-\Delta S_M^{max}$ increases continuously with the increasing field, and a faster increase of $-\Delta S_M^{max}$ with increasing field was observed above 2 T. I.e. the MCE could attain higher value in higher magnetic fields. The MCE parameters for $Dy_{0.9}Tm_{0.1}Ni_2B_2C$ are comparable with those of the potential promising magnetic refrigerant materials in the similar temperature region.[104]

### 4.2. Large magnetic entropy change in $Dy_{1-x}Ho_xNi_2B_2C$ (x = 0 - 1) superconductors

We have also investigated the superconductivity, magnetic properties and MCE in $Dy_{1-x}Ho_xNi_2B_2C$ (x = 0 - 1) superconductors. The superconducting transition temperature $T_{SC}$ and the magnetic ordering temperature $T_M$ are determined to be 6.4, 6.4, 6.2, 8.1 and 8.2 K and to be 10, 8.5, 8, 6 and 5 K for x = 0, 03, 0.5, 0.7 and 1, respectively.[105, 106] A set of selected magnetic isothermals on increasing and decreasing field for $Dy_{1-x}Ho_xNi_2B_2C$ (x = 0, 0.3, 0.5, 0.7 and 1) are measured in the temperature range from 2 to 36 K up to 7 T. Although some hysteresis of M(H) can be observed at low temperatures, it gradually became smaller with increasing temperature, and almost disappeared above 5 K. All the samples show similar behaviors. Several isotherms with increasing field for x = 0.5 are presented in Fig. 18(a). To understand the order of magnetic transition in $Dy_{1-x}Ho_xNi_2B_2C$, the measured M-H isotherms were converted in to H/M versus $M^2$ plots. A clear negative slope at low temperature and low magnetic field region can be observed for all the compounds, the Arrott plots of x = 0.5 are shown in Fig. 18(b) for an example, indicating the first order magnetic phase transition for $Dy_{1-x}Ho_xNi_2B_2C$ (x = 0, 0.3, 0.5, 0.7 and 1) compounds.[106]

The magnetic entropy change -$\Delta S_M$ was calculated from magnetization isotherm M (H, T) curves. The resulted temperature dependence of -$\Delta S_M$ is shown in Fig 19 for x = 0, 0.3, 0.7 and 1 and Fig. 20 (b) for x = 0.5 in $Dy_{1-x}Ho_xNi_2B_2C$. The magnetic entropy change for $Dy_{0.5}Ho_{0.5}Ni_2B_2C$ were also calculated from the field and temperature dependence of heat capacity C (T, H) [(as shown in Fig. 20 (a)] using $\Delta S_M(T,\Delta H) = \int_0^T [C(T,H_1) - C(T,H_0)/T] dT$, and the results are also shown in Fig. 20 (b). The results of entropy changes calculated from $M$ $(T, H)$ and $C$ $(T, H)$ are consistent with each other. For low magnetic field changes, $-\Delta S_M$ is negative (inverse MCE) below the transition temperature, and changes to positive with increasing temperature which can be understood in terms of the FM-AFM phase coexistence and the variation in the ratio of these phases under different magnetic fields. For the magnetic field changes of 0-5 and 0-7 T, a positive $-\Delta S_M$ (i.e. conventional MCE) with a broad peak can be observed. For the magnetic field changes of 0-2, 0-5 and 0-7 T, the maximum values of magnetic entropy change ($-\Delta S_M^{max}$) are evaluated to be 4.5, 17.1 and 22.1 J/kg K for $x = 0$; to be 7.3, 20.2 and 23.3 J/kg K for $x = 0.3$; to be 5.9, 18.5 and 22.4 J/kg K for $x = 0.5$; to be 6.3, 18.7 and 22.9 J/kg K for $x = 0.7$; and to be 7.3, 19.2 and 22.3 J/kg K for $x = 1$, respectively.[106] The large magnetic entropy changes in $Dy_{1-x}Ho_xNi_2B_2C$ are believed to be related to the field induced first-order magnetic transition from AFM to FM. The values of $RCP$ are all keeping at same high values in the range of 45-62, 243-290 and 414-510 J/kg for the magnetic field changes of 0-2, 0-5 and 0-7 T, respectively. The values of $-\Delta S_M^{max}$ and $RCP$ for $Dy_{1-x}Ho_xNi_2B_2C$ ($x = 0, 0.3, 0.5, 0.7$ and 1) are comparable with some of potential magnetic refrigerant materials reported in the same temperature range which indicate that the $Dy_{1-x}Ho_xNi_2B_2C$ compounds could be promising candidates for low temperature magnetic refrigeration.[106]

### 4.3. Ni site doping effect on MCE for $RE$Ni$_{2-x}$A$_x$B$_2$C ($RE$ = Dy and Er)

The magnetic properties and MCE in $DyNi_{2-x}A_xB_2C$ ($A$ = Co and Cr, $x$ = 0.1, and 0.2) compounds were also investigated. No superconductivity can be observed above 2 K, and the magnetic transition temperature $T_M$ are determined to be 9.8, 8.4, 8.0, 9.2 and 8.8 K for $x = 0$, 0.1 (Co), 0.2 (Co), 0.1 (Cr) and 0.2 (Cr) in $DyNi_{2-x}A_xB_2C$ system, respectively. Figure 21 shows the magnetic isothermals on increasing (open symbols) and decreasing (filled symbols) field for $DyNi_{2-x}A_xB_2C$ ($A$ = Co and Cr, $x$ = 0.1 and 0.2) at 2 K, the hysteresis is ignorable for all the present DyNi$_{2-x}$A$_x$B$_2$C which is beneficial for application. I. e. similar to that of Dy site Tm substitution, [104] Ni site Co or Cr substitution also can effectively reduce (even eliminate) the magnetic hysteresis of DyNi$_2$B$_2$C.[107]

To evaluate the MCE in DyNi$_{2-x}$A$_x$B$_2$C (A = Co and Cr, x = 0.1 and 0.2), a set of magnetic isothermals are measured. All the samples show similar behaviors. Several isotherms of DyNi$_{1.8}$A$_{0.2}$B$_2$C for A = Co and Cr are presented in Fig. 22(a) and (b), and the corresponding Arrott-plot are presented in Fig. 22(c) and (d), respectively. Ni site Co or Cr substitution lower the magnetic transition temperature $T_M$, and reduce the magnetic hysteresis of DyNi$_2$B$_2$C. The magnetic properties and magnetocaloric effect (MCE) in DyNi$_{2-x}$A$_x$B$_2$C (A = Co and Cr, x = 0.1, and 0.2) compounds have been studied. Ni site Co or Cr substitution lower the magnetic transition temperature $T_M$, and reduce the magnetic hysteresis of DyNi$_2$B$_2$C. A clear negative slope at low temperature and low magnetic field region can be observed in Fig. 22 (c) and (d), indicating the first order magnetic phase transition for DyNi$_{2-x}$A$_x$B$_2$C (A = Co and Cr, x = 0.1 and 0.2) compounds.

The temperature and field dependences -$\Delta S_M$ for DyNi$_{2-x}$A$_x$B$_2$C (A = Co and Cr, x = 0.1 and 0.2) were calculated based on the M(H, T) curves (as shown in Fig. 23). For the magnetic field changes of 0-1 and 0-2 T, -$\Delta S_M$ is negative (inverse MCE) below the transition temperature, and changes to positive with increasing temperature, for the magnetic field changes of 0-5 and 0-7 T, a positive -$\Delta S_M$ (i.e., conventional MCE) can be observed. This behaviour is similar to that of Dy$_{1-x}$Ho$_x$Ni$_2$B$_2$C compounds[106] which can be understood in terms of the FM and AFM phase coexistence and the variation in the ratio of these phases under different magnetic fields. For the magnetic field change of 0-2 T, the minimum and maximum values of -$\Delta S_M$ are evaluated to be -1.4 and 4.7 J/kg K for DyNi$_{1.9}$Co$_{0.1}$B$_2$C, to be -1.9 and 2.3 J/kg K for DyNi$_{1.8}$Co$_{0.2}$B$_2$C, to be -1.4 and 4.9 J/kg K for DyNi$_{1.9}$Cr$_{0.1}$B$_2$C, and to be -0.7 and 4.6 J/kg K for DyNi$_{1.8}$Cr$_{0.2}$B$_2$C, respectively. The values of $-\Delta S_M^{max}$ are evaluated to be 16.3, 10.2, 16.1 and 13.7 J/kg K; and to be 19.3, 13.7, 19.1 and 15.8 J/kg K for $x$ = 0.1 (Co), 0.2 (Co), 0.1 (Cr) and 0.2 (Cr) in DyNi$_{2-x}$A$_x$B$_2$C for the magnetic field changes of 0-5 and 0-7 T, respectively.[107]

Zhang and Yang [108] have investigated the magnetic properties and magnetocaloric effect (MCE) in ErNi$_{2-x}$Fe$_x$B$_2$C ($x$ = 0, 0.1 and 0.2) compounds. Similar to those of DyNi$_{2-x}$A$_x$B$_2$C ($A$ = Co and Cr, $x$ = 0.1 and 0.2), substitution of Fe for Ni lowered the magnetic transition temperature $T_M$, and reduced the magnetic hysteresis of ErNi$_2$B$_2$C. The inverse MCE under low magnetic field and at low temperatures is attributed to the nature of antiferromagnetic state for the present ErNi$_{2-x}$Fe$_x$B$_2$C compounds. A normal MCE was observed under higher magnetic field changes, which is related to a field-induced first order metamagnetic transition from AFM to FM state. The maximum values of magnetic entropy change $-\Delta S_M^{max}$ are 14.5, 12.7 and 10.6 J/kg K for the magnetic field change of 0-7 T for $x$ = 0, 0.1 and 0.2 in ErNi$_{2-x}$Fe$_x$B$_2$C, respectively.[108]

## 5. Summary

Investigations on MCE are of great interest for fundamental properties and technological application. Our recent progress of the magnetic properties and MCE as well as its origin and potential application for magnetic refrigeration in some rare earth based intermetallic compounds are reviewed, and some of them are possess excellent MCE properties. The main achievements are: 1) The TmZn compound exhibits a ferromagnetic state below the Curie temperature of $T_C$ = 8.4 K, and processes a field-induced metamagnetic phase transition around and above $T_C$, resulting in a giant reversible MCE. Particularly, the values of $-\Delta S_M^{max}$ reach 11.8 and 19.6 J/kg K for a low magnetic field changes of 0-1 and 0-2 T, respectively, indicating that TmZn could be a promising candidate for low temperature and low field magnetic refrigeration. 2) A single phased Eu$_4$PdMg compounds was synthesised by induction melting of the elements in a sealed tantalum tube in a water-cooled sample chamber. Eu$_4$PdMg has a very magnetic field sensitive magnetic phase transition, resulting in a reversible, table-like MCE over a broad temperature range. Additionally, the *RCP* values are obviously larger than those of some potential magnetic refrigerant materials at similar temperature region, making Eu$_4$PdMg attractive for magnetic refrigeration from 20 to 160 K. 3) The GdCd$_{1-x}$Ru$_x$ solid solutions with *x* = 0.1, 0.15 and 0.2 have been successfully synthesized and the magnetism and MCE have been investigated. The samples undergo a second order magnetic phase transition from a paramagnetic to a ferromagnetic state at Curie temperatures of $T_C \sim$ 160, 108 and 73 K for *x* = 0.1, 0.15 and 0.2, respectively. A large reversible MCE with a wide temperature range in GdCd$_{1-x}$Ru$_x$ solid solutions was observed which make the GdCd$_{1-x}$Ru$_x$ solid solutions considerable for active magnetic-refrigeration. 4) The Tm substitution effectively reduces the thermal hysteresis of DyNi$_2$B$_2$C, therefore, a giant reversible MCE has been observed in antiferromagnetic superconductor Dy$_{0.9}$Tm$_{0.1}$Ni$_2$B$_2$C which is related to a field-induced first order metamagnetic transition from AFM to FM state. 5) The magnetic transition temperature $T_M$ as well as the temperature of -$\Delta S_M^{max}$ in antiferromagnetic superconductors Dy$_{1-x}$Ho$_x$Ni$_2$B$_2$C gradually shifts to low temperature with increasing *x*, while the values of -$\Delta S_M^{max}$ remain at almost the same high value which make the Dy$_{1-x}$Ho$_x$Ni$_2$B$_2$C compounds attractive for active magnetic-refrigeration for low temperature magnetic refrigeration. The present results may provide some valuable information for searching proper magnetic materials for low temperature magnetic refrigeration.

**Figure captions**

Fig. 1. (Colour online) The total entropy *S* as function of magnetic field *H* and temperature *T* schematically illustrating the definition of the isothermal magnetic entropy change $\Delta S_M$ and the adiabatic temperature change $\Delta T_{ad}$.

Fig. 2. (Colour online) An example for the defination of the relative cooling power (*RCP*) and refrigerant capacity (*RC*) for $GdRu_{0.2}Cd_{0.8}$. Rectangular area is *RCP* and the area full with parallel lines is *RC*.

Fig. 3. Temperature dependence of the zero field cooling (ZFC) and field cooling (FC) magnetization (*M*) for TmZn under various magnetic fields (*H*) up to 1 T.[57]

Fig. 4. Magnetic field dependence of the magnetization (increasing field only) for TmZn at some selected temperatures.[57]

Fig. 5. The plots of *H*/*M* versus $M^2$ for TmZn at some selected temperatures.[57]

Fig. 6. (Colour online) Temperature dependence of magnetic entropy change $-\Delta S_M$ for TmZn with the magnetic field changes up to 0-7 T.[57]

Fig. 7. (Colour online) Temperature dependence of adiabatic temperature change $\Delta T_{ad}$ for TmZn compound with the magnetic field changes up to 0-7 T.[57]

Fig. 8. The magnetic entropy change $-\Delta S_M$ as a function of temperature for various magnetic field changes up to 0-7 T for HoZn.[58]

Fig. 9. (Colour online) (a): Temperature dependence of zero field cooling (ZFC) and field cooling (FC) magnetization *M* (left scale) as well as $dM_{FC}/dT$ (right scale) for $Eu_4PdMg$ in an external magnetic field *H* = 0.1 T. (b): Temperature dependence of FC magnetization *M* (left scale) and $dM_{FC}/dT$ (right scale) for $Eu_4PdMg$ under *H* = 1 T.[66]

Fig. 10. Temperature dependences of the magnetization for $Eu_4PdMg$ under various magnetic fields up to 7 T.[66]

Fig. 11. Temperature dependence of magnetic entropy change $-\Delta S_M$ for $Eu_4PdMg$ with the magnetic field changes up to 0-7 T.[66]

Fig. 12. Temperature dependence of magnetic entropy change $-\Delta S_M$ with the magnetic field

changes up to 0-7 T for Ho$_4$PtMg (a), Er$_4$PdMg (b) and Er$_4$PtMg (c), respectively. [67, 68]

Fig. 13. Temperature dependence of adiabatic temperature change $\Delta T_{ad}$ with the magnetic field changes up to 0-7 T for Ho$_4$PtMg (a), Er$_4$PdMg (b) and Er$_4$PtMg (c), respectively. [67, 68]

Fig. 14. (Colour online) Temperature dependence of zero field cooling (ZFC) and field cooling (FC) magnetization $M$ for GdCd$_{1-x}$Ru$_x$ ($x$ = 0.1, 0.15, and 0.2) under a low magnetic field of $H$ = 0.1 T. [73]

Fig. 15. Temperature dependence of magnetic entropy change -$\Delta S_M$ with various magnetic field changes for GdCd$_{1-x}$Ru$_x$ ($x$ = 0.1, 0.15, and 0.2). The inset of (c) shows the maximum magnetic entropy change -$\Delta S_M^{max}$ as a function of magnetic field change $\Delta H^{2/3}$. [73]

Fig. 16. Magnetic isothermals on increasing (open symbols) and decreasing (filled symbols) field for Dy$_{0.9}$Tm$_{0.1}$Ni$_2$B$_2$C at 2 and 3 K as well for DyNi$_2$B$_2$C at 2 K. [104]

Fig. 17. Temperature dependence of magnetic entropy change -$\Delta S_M$ for Dy$_{0.9}$Tm$_{0.1}$Ni$_2$B$_2$C. The inset shows the maximum values of magnetic entropy change -$\Delta S_M^{max}$ as a function of magnetic field changes. [104]

Fig. 18. (a) Magnetic field dependence of the magnetization for Dy$_{0.5}$Ho$_{0.5}$Ni$_2$B$_2$C at some selected temperatures. (b): The plots of $H/M$ versus $M^2$ for Dy$_{0.5}$Ho$_{0.5}$Ni$_2$B$_2$C at some selected temperatures. [105]

Fig. 19. Temperature dependence of magnetic entropy change -$\Delta S_M$ calculated from $M(T, H)$ for Dy$_{1-x}$Ho$_x$Ni$_2$B$_2$C ($x$ = 0, 0.3, 0.7 and 1). [105]

Fig. 20. (Colour online) (a) Temperature dependence of heat capacity under the magnetic field of 0 T, 2 T and 5 T for Dy$_{0.5}$Ho$_{0.5}$Ni$_2$B$_2$C. (b) Temperature dependence of magnetic entropy change -$\Delta S_M$ for Dy$_{0.5}$Ho$_{0.5}$Ni$_2$B$_2$C calculated from $C(T, H)$ and $M(T, H)$, respectively. [105]

Fig. 21. (Colour online) Magnetic isothermals on increasing (open symbols) and decreasing (filled symbols) field at 2 K for DyNi$_{2-x}$A$_x$B$_2$C ($A$ = Co and Cr, $x$ = 0.1 and 0.2). [107]

Fig. 22. Magnetic isothermals at some selected temperature for DyNi$_{1.8}$Co$_{0.2}$B$_2$C (a) and DyNi$_{1.8}$Cr$_{0.2}$B$_2$C (b). The curves of $H/M$ versus $M^2$ for DyNi$_{1.8}$Co$_{0.2}$B$_2$C (c) and DyNi$_{1.8}$Cr$_{0.2}$B$_2$C (d) at some selected temperatures. [107]

Fig. 23. Temperature dependence of magnetic entropy change -$\Delta S_M$ for DyNi$_{2-x}$A$_x$B$_2$C ($A$ = Co and Cr, $x$ = 0.1 and 0.2). [107]

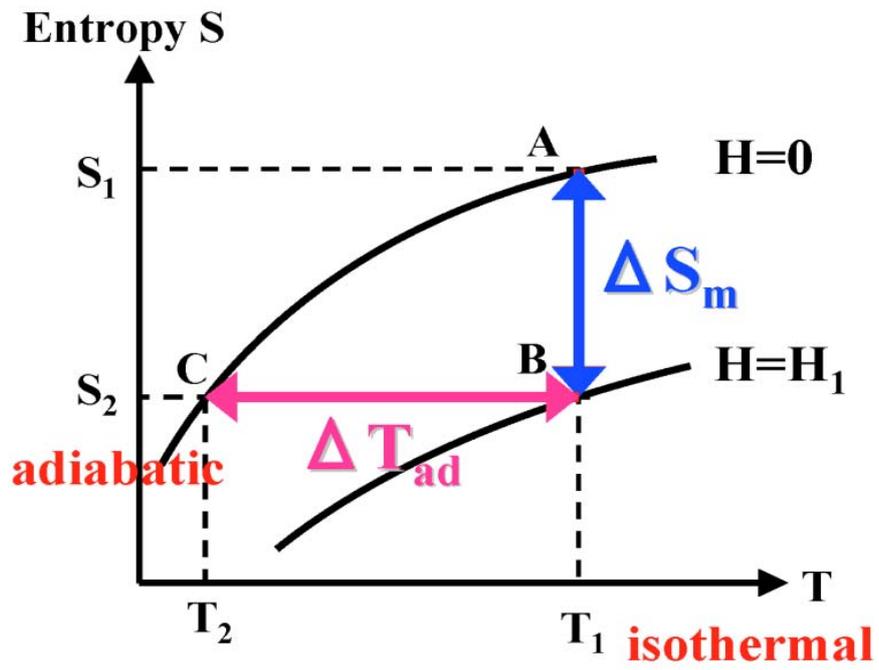

Fig. 1.

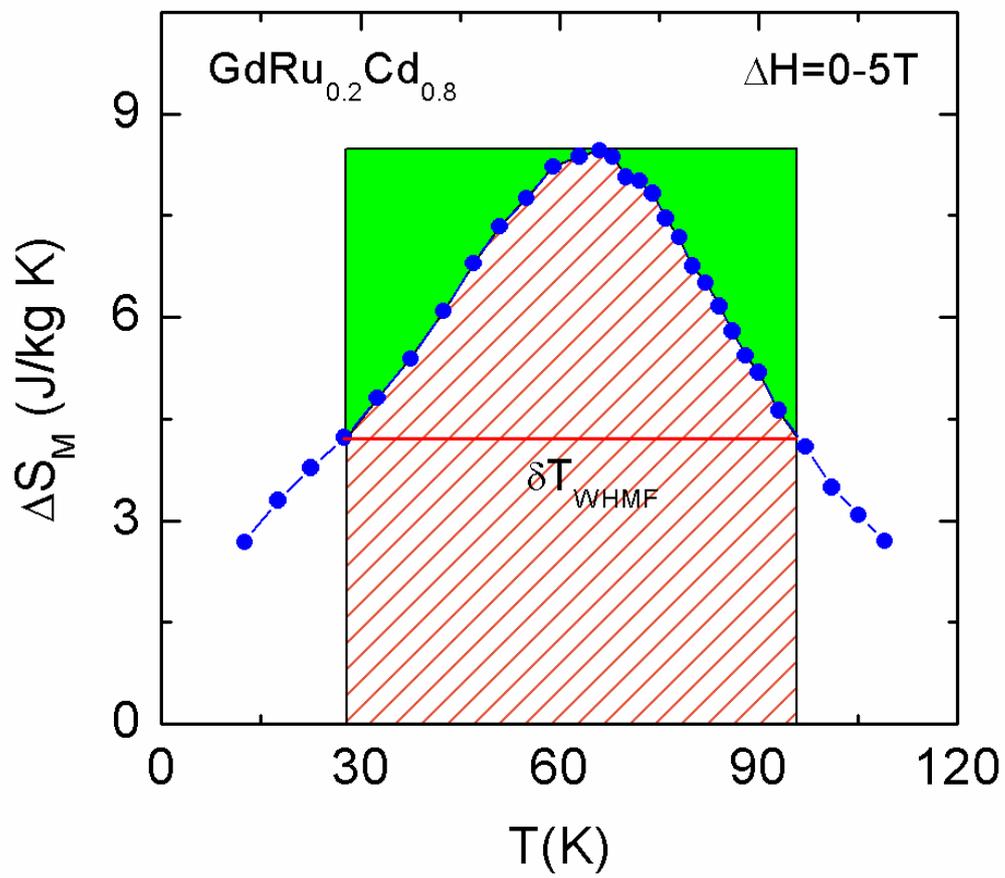

Fig. 2.

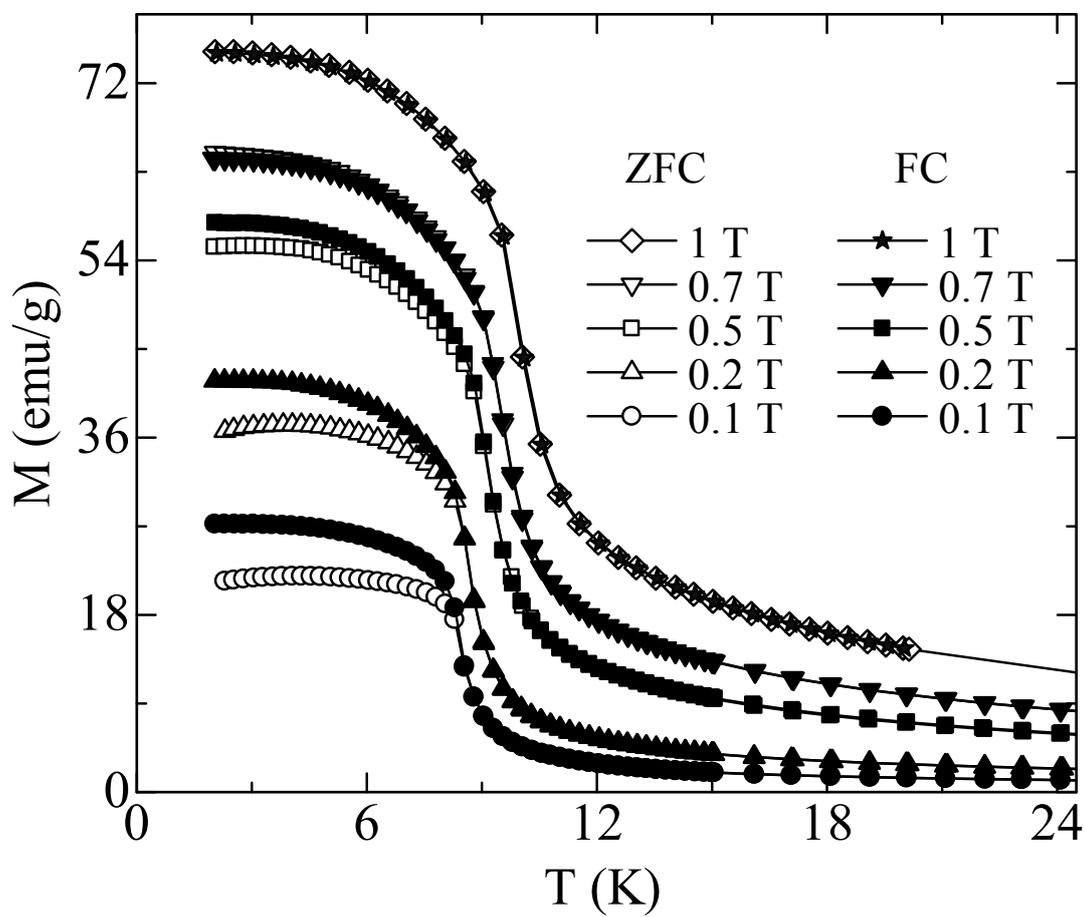

Fig. 3.

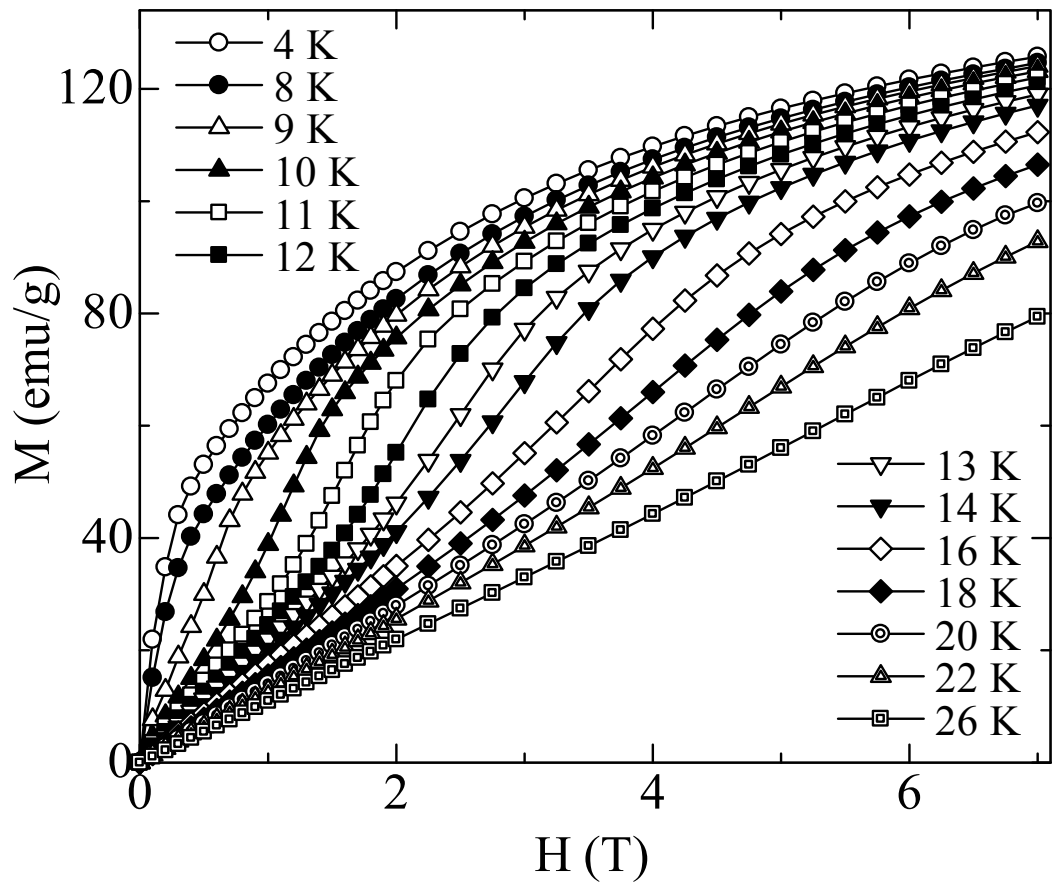

Fig. 4.

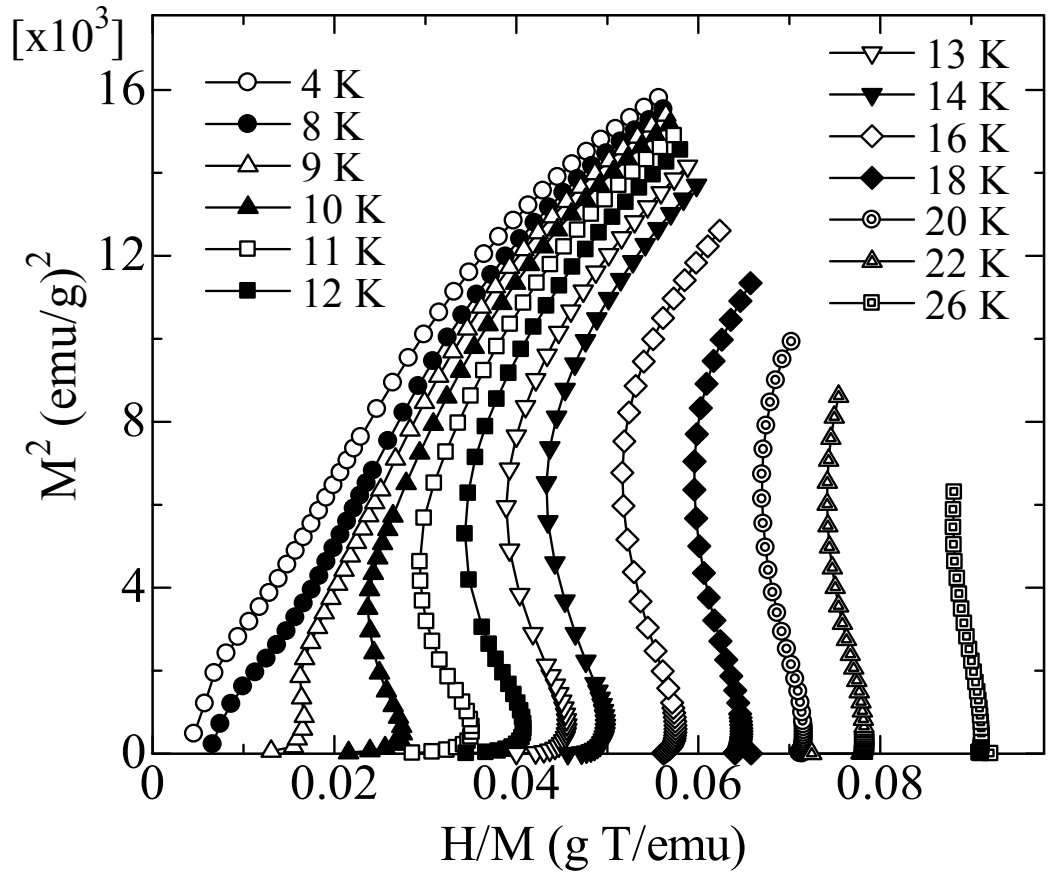

Fig. 5.

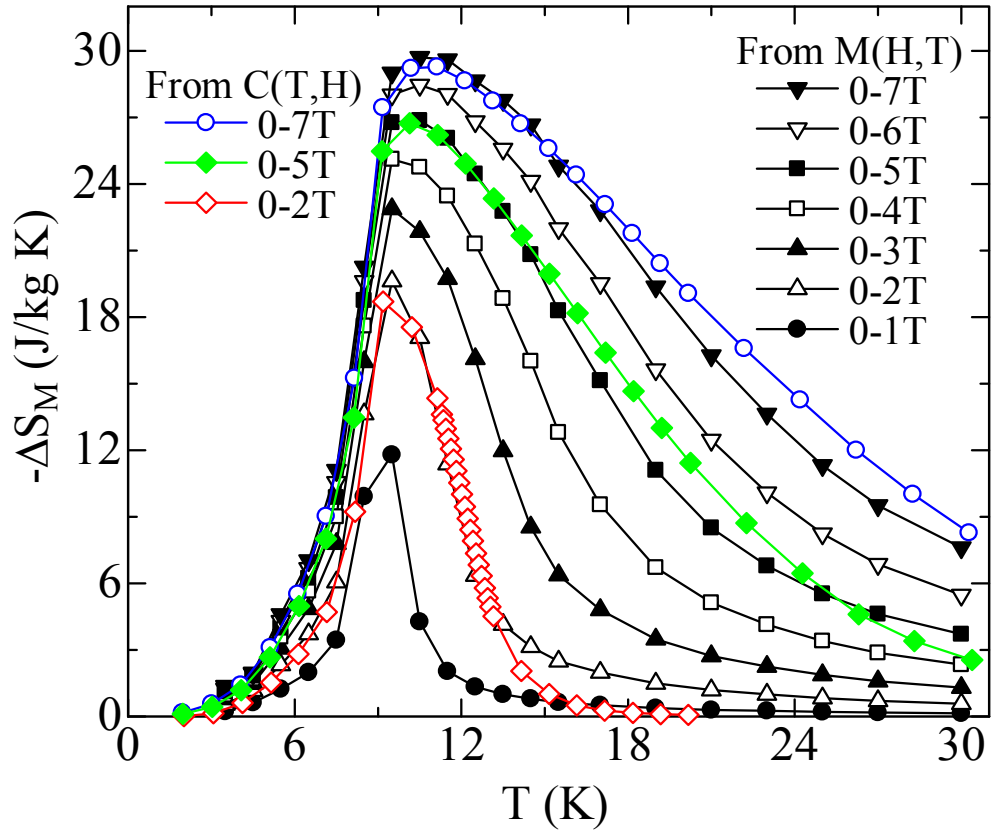

Fig. 6.

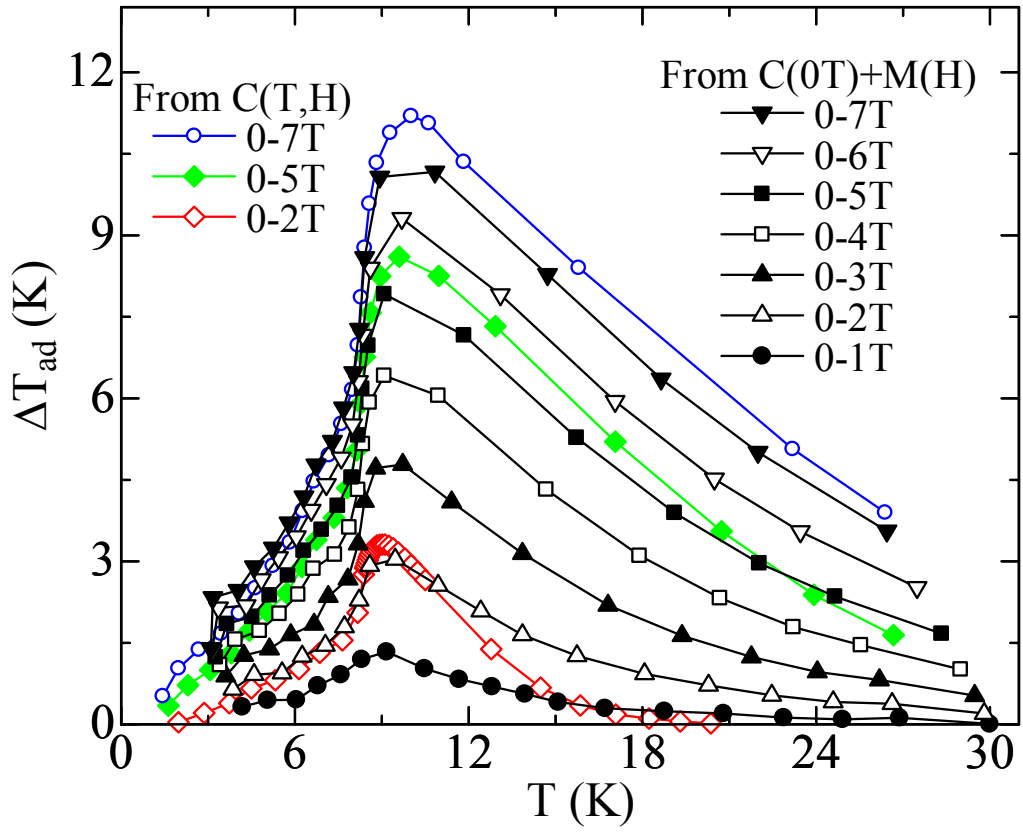

Fig. 7.

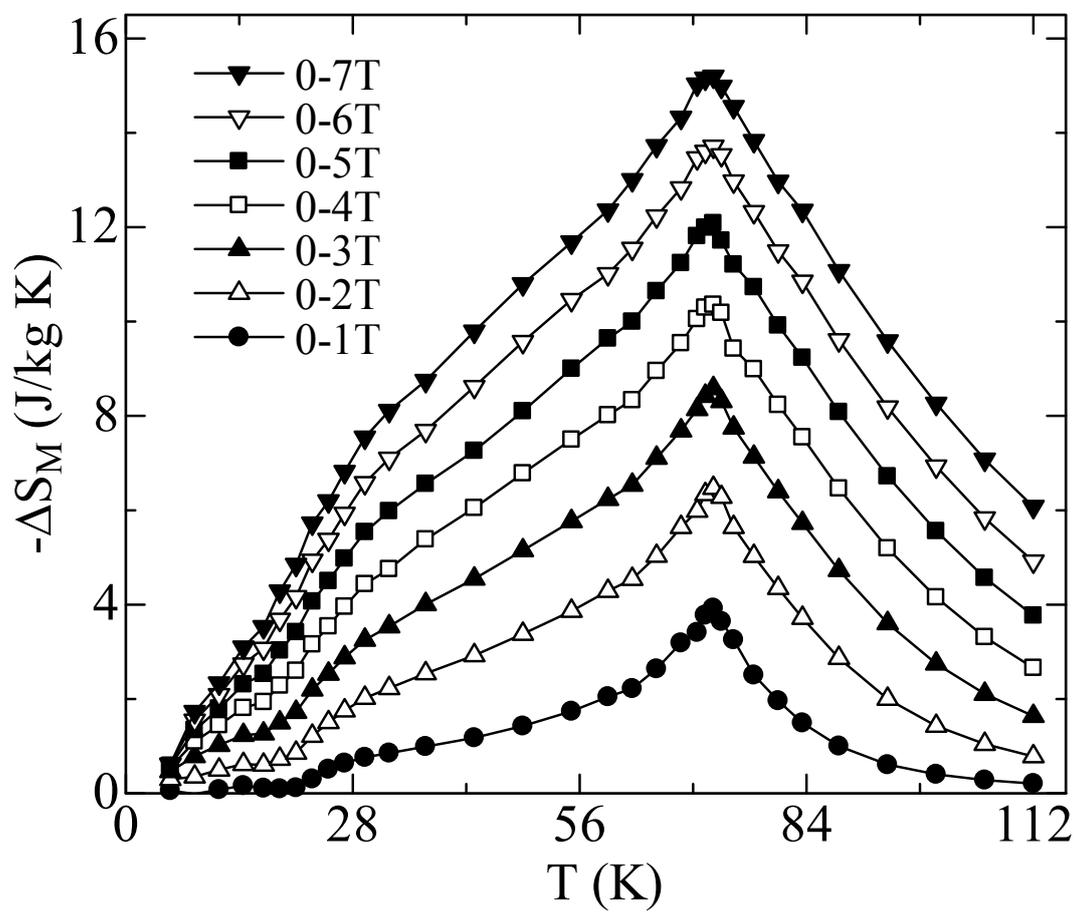

Fig. 8.

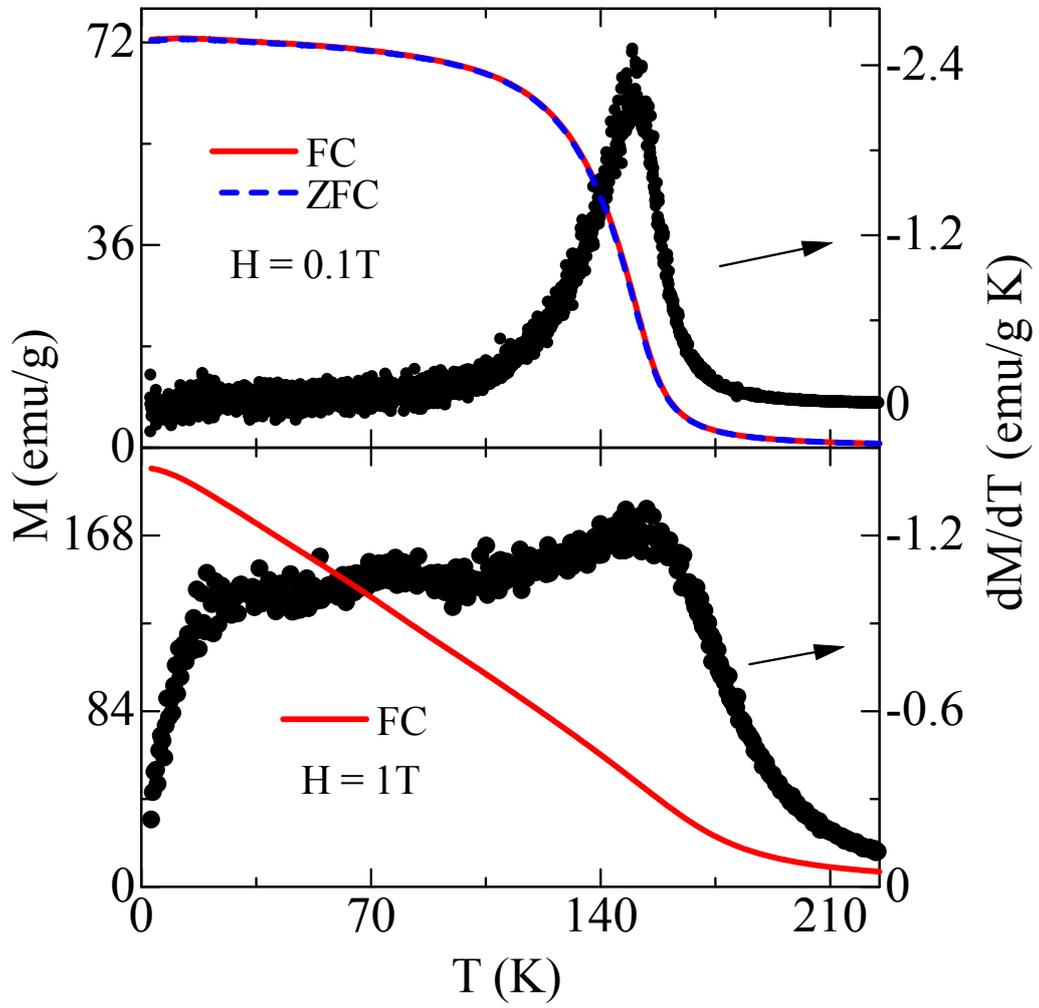

Fig. 9.

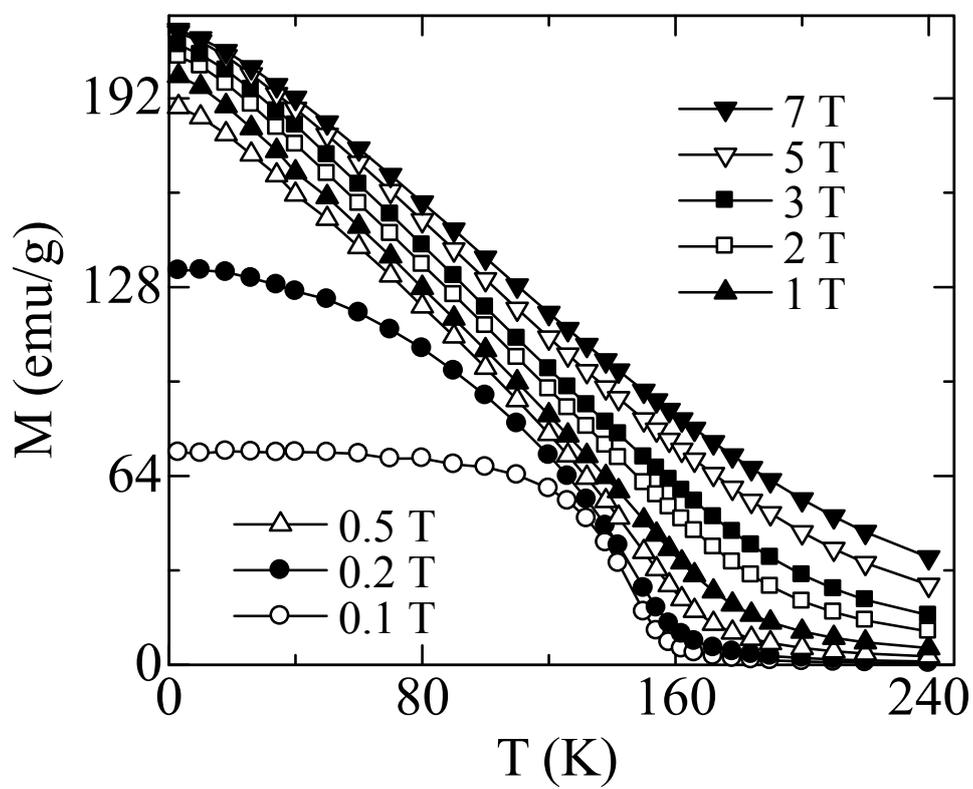

Fig. 10.

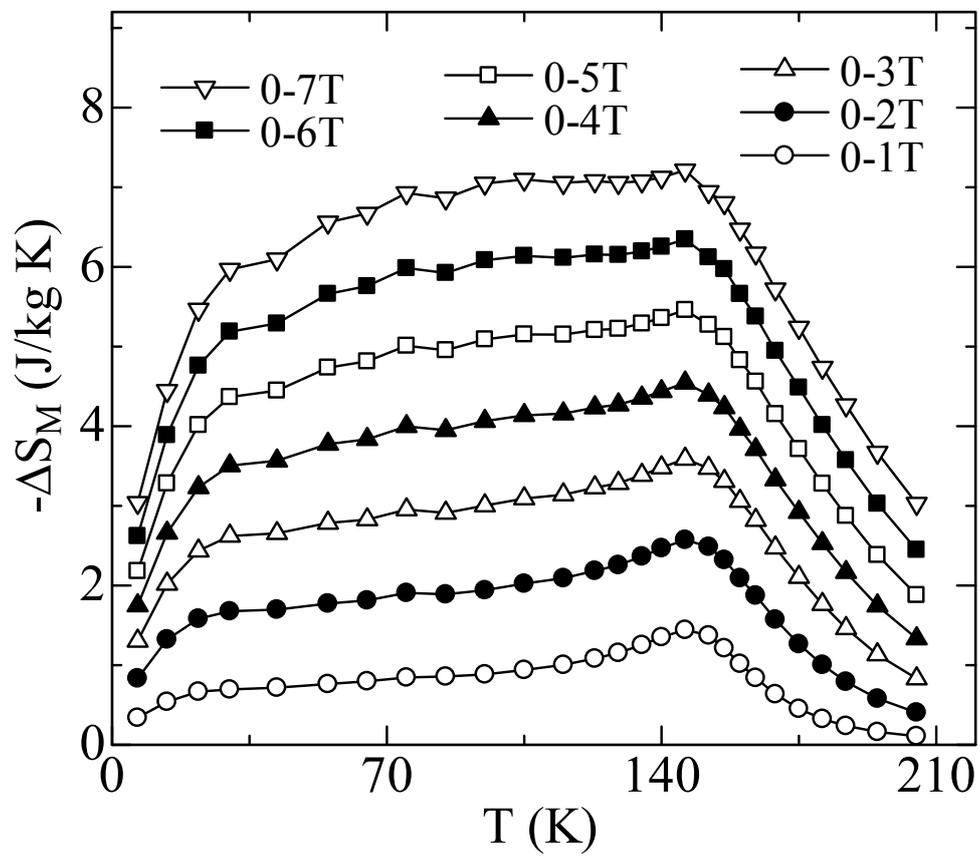

Fig. 11.

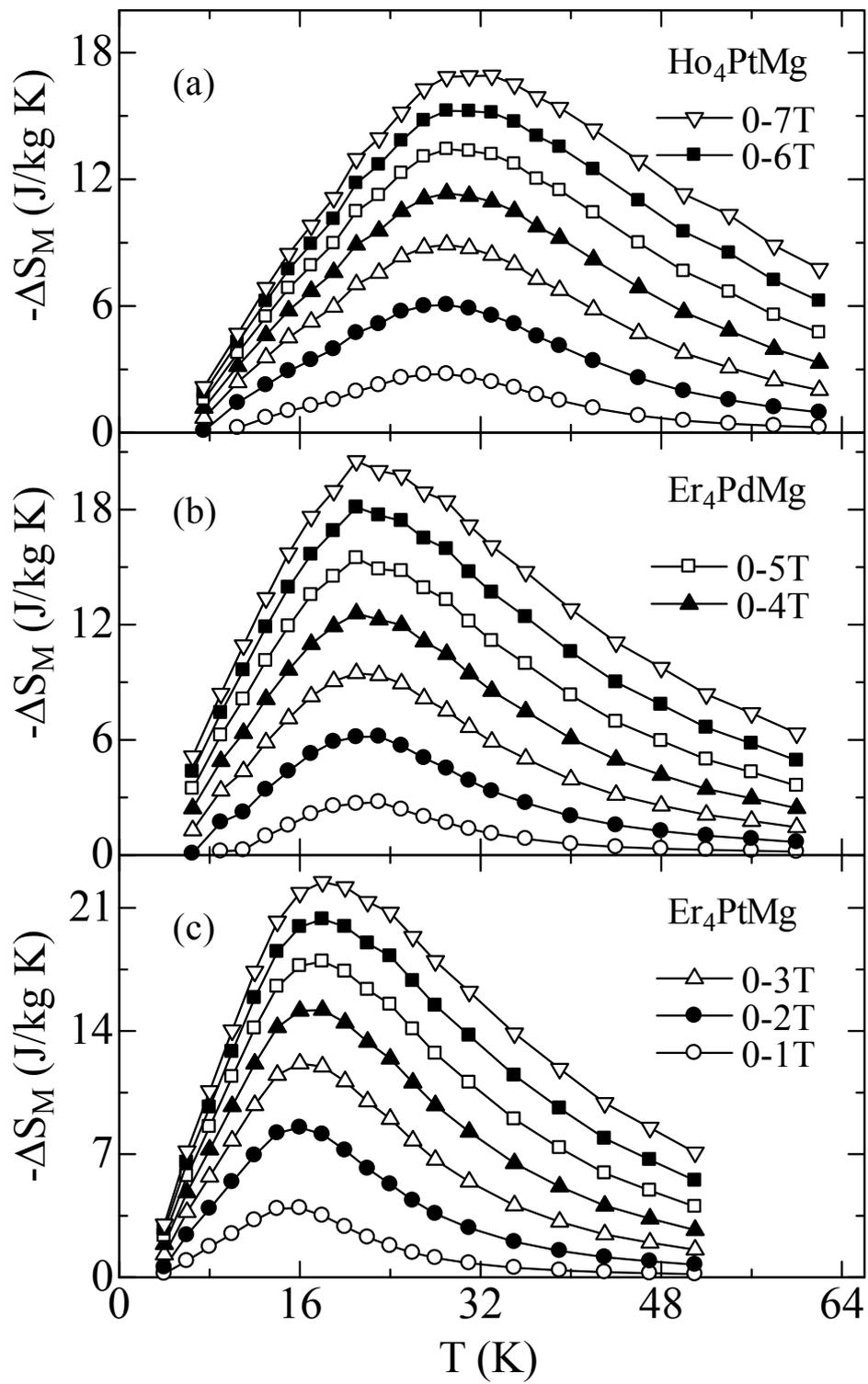

Fig. 12.

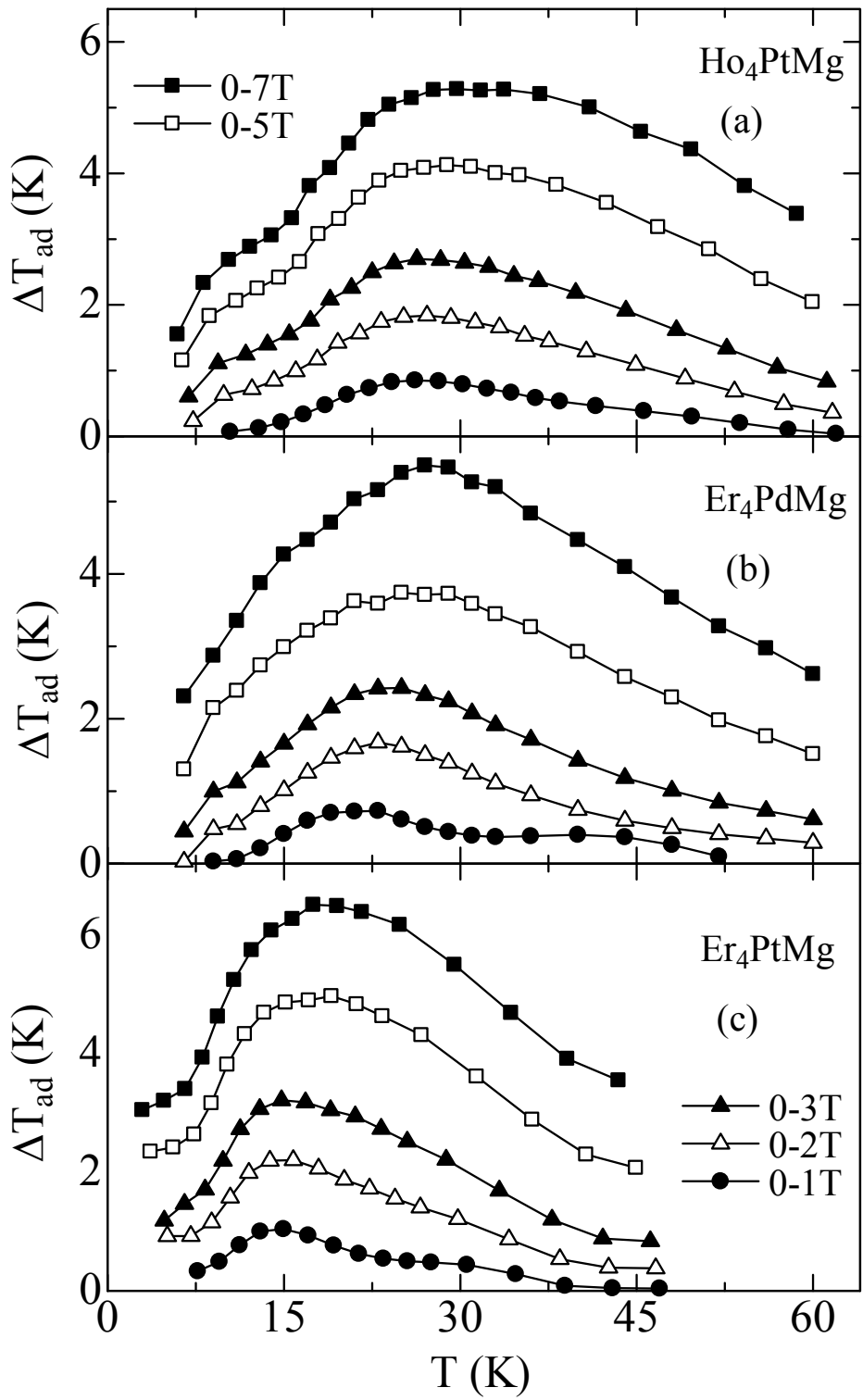

Fig. 13.

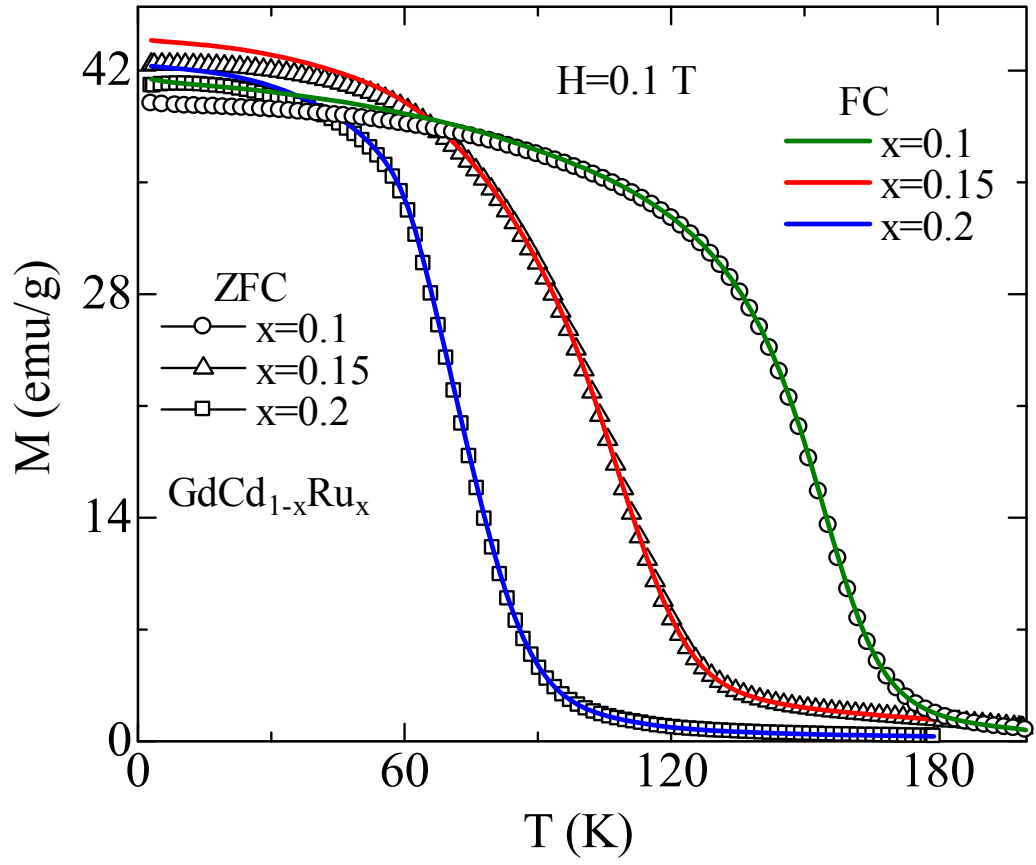

Fig. 14.

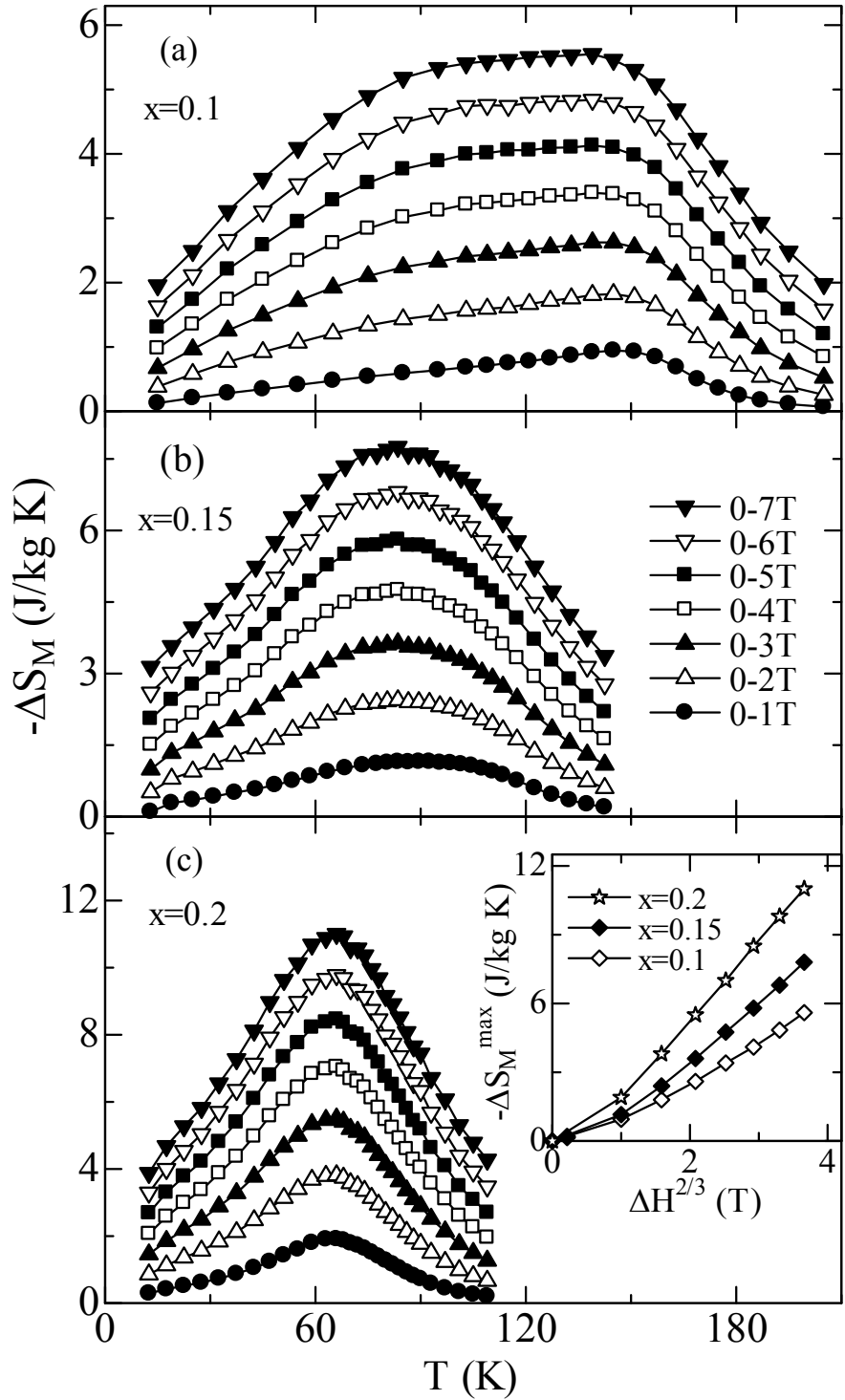

Fig. 15.

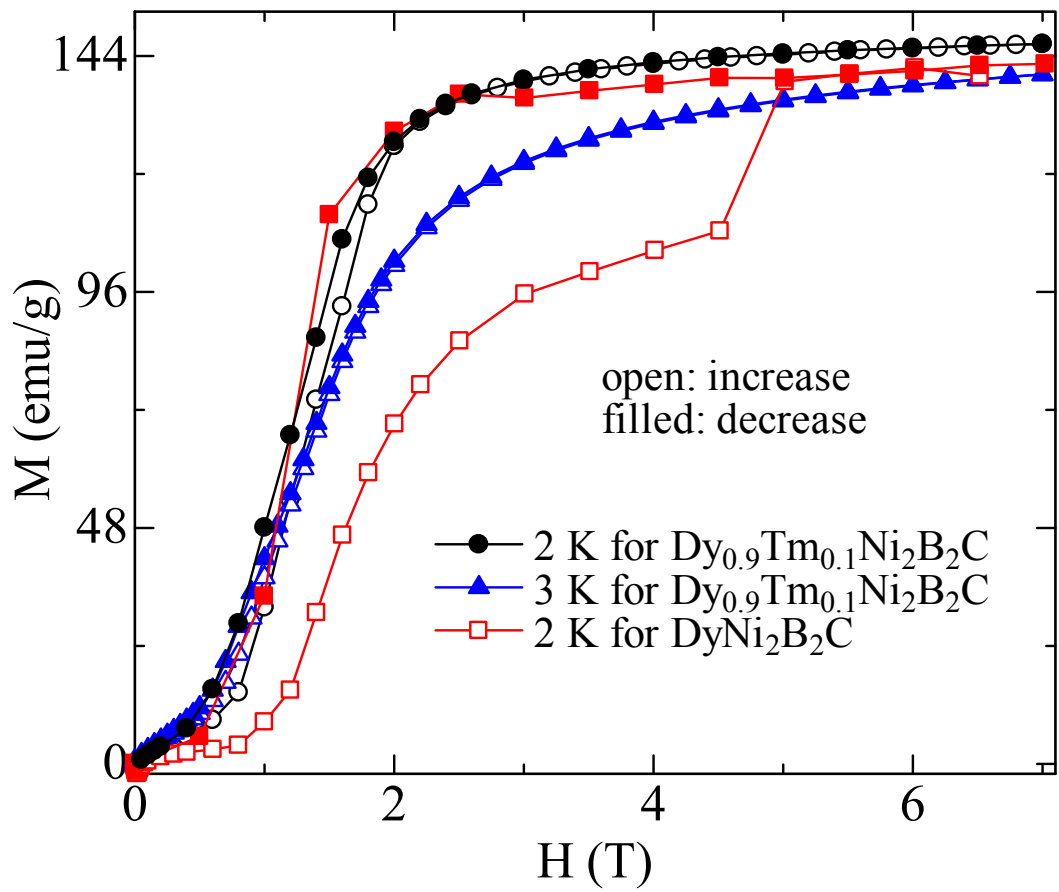

Fig. 16.

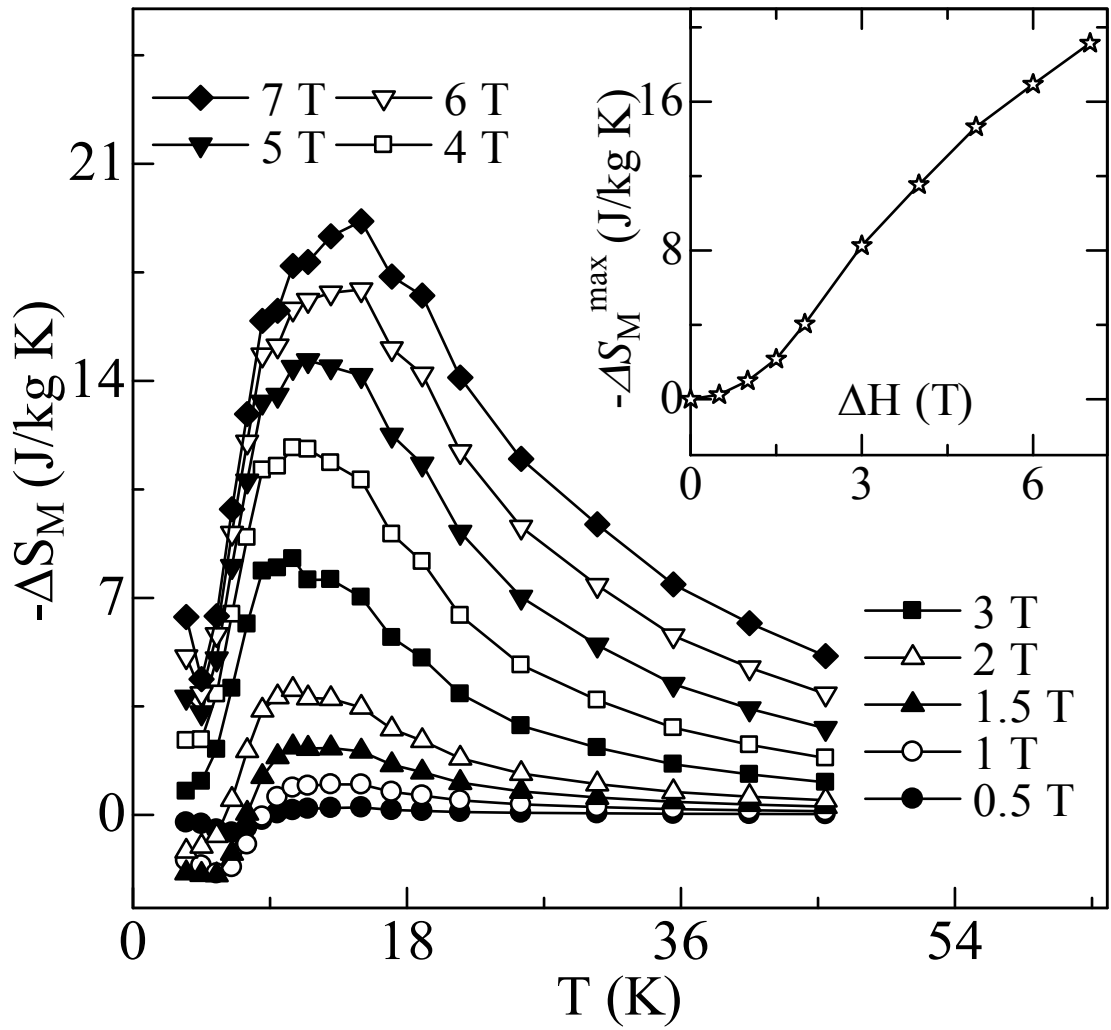

Fig. 17.

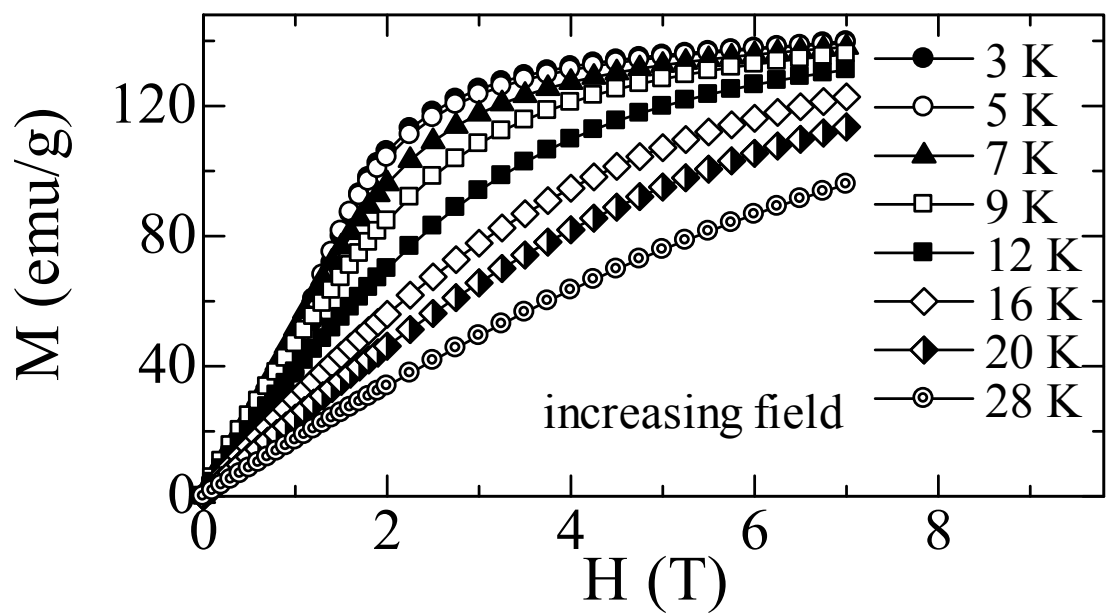

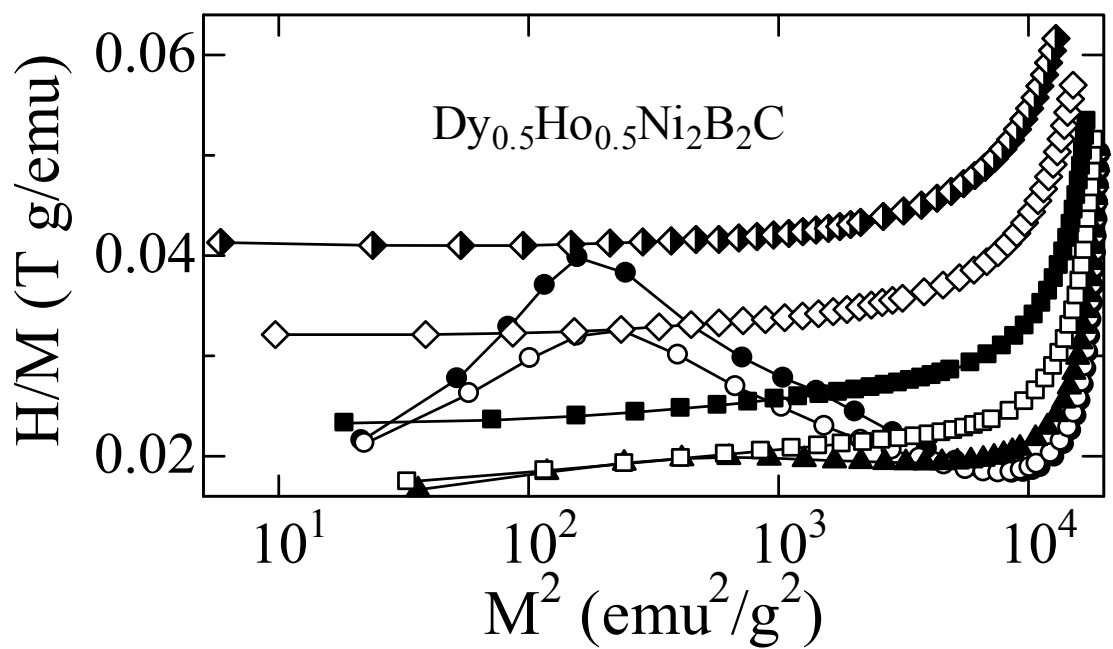

Fig. 18.

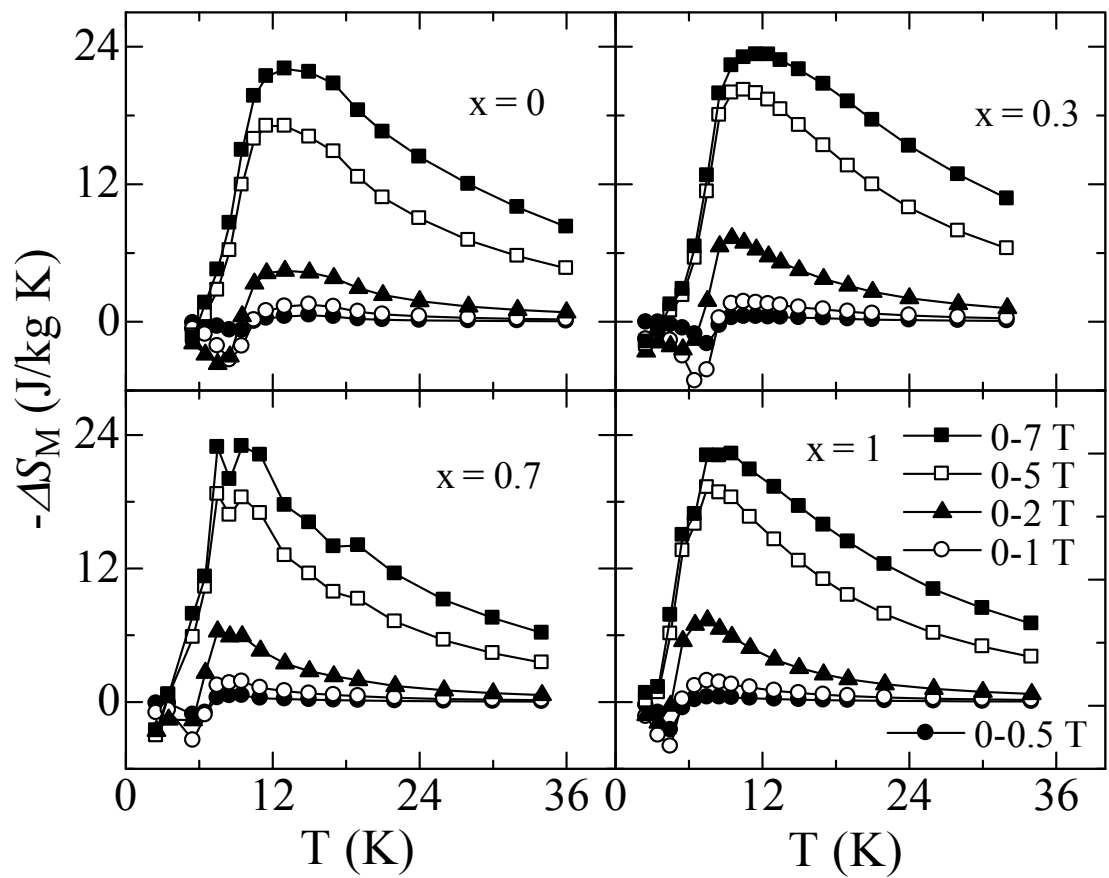

Fig. 19.

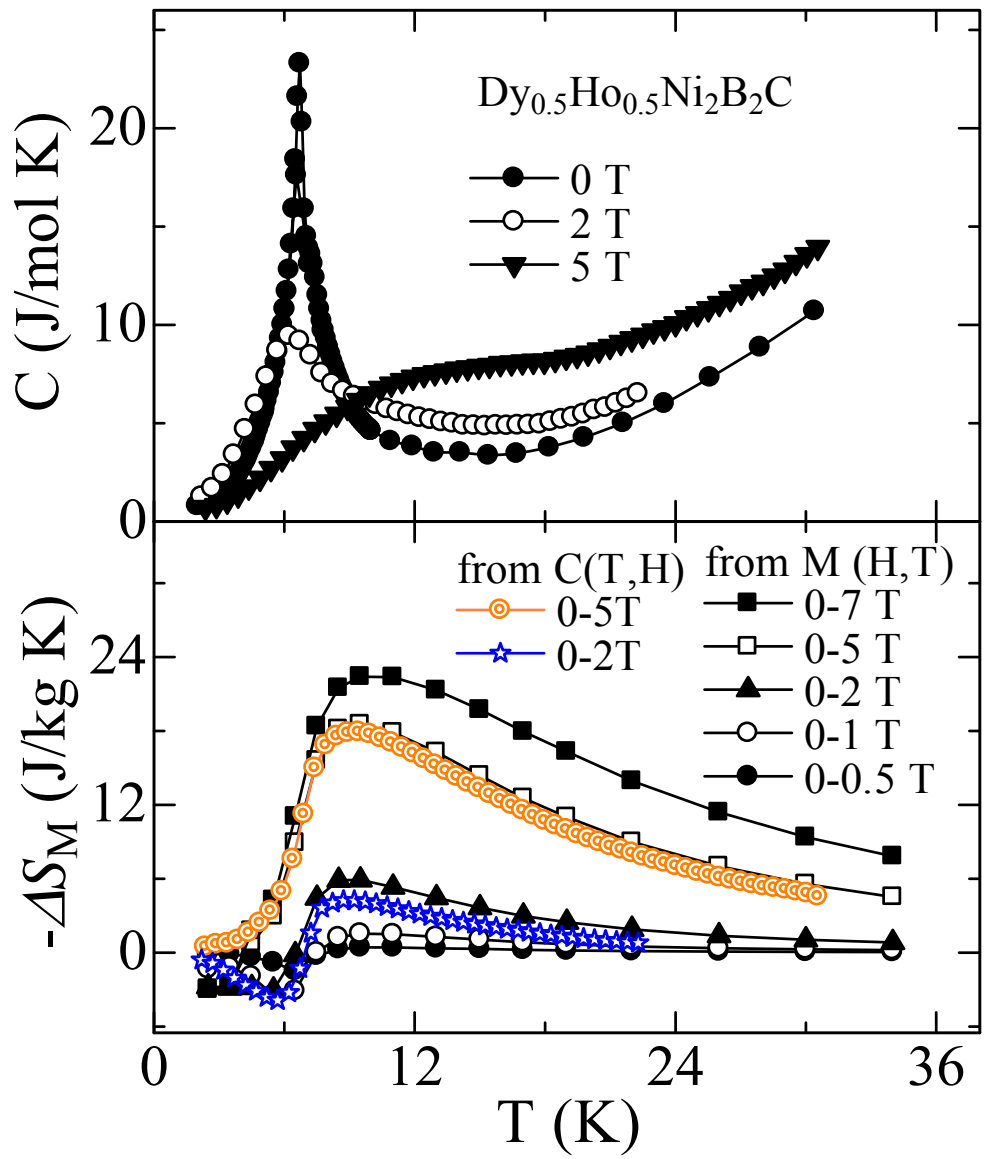

Fig. 20.

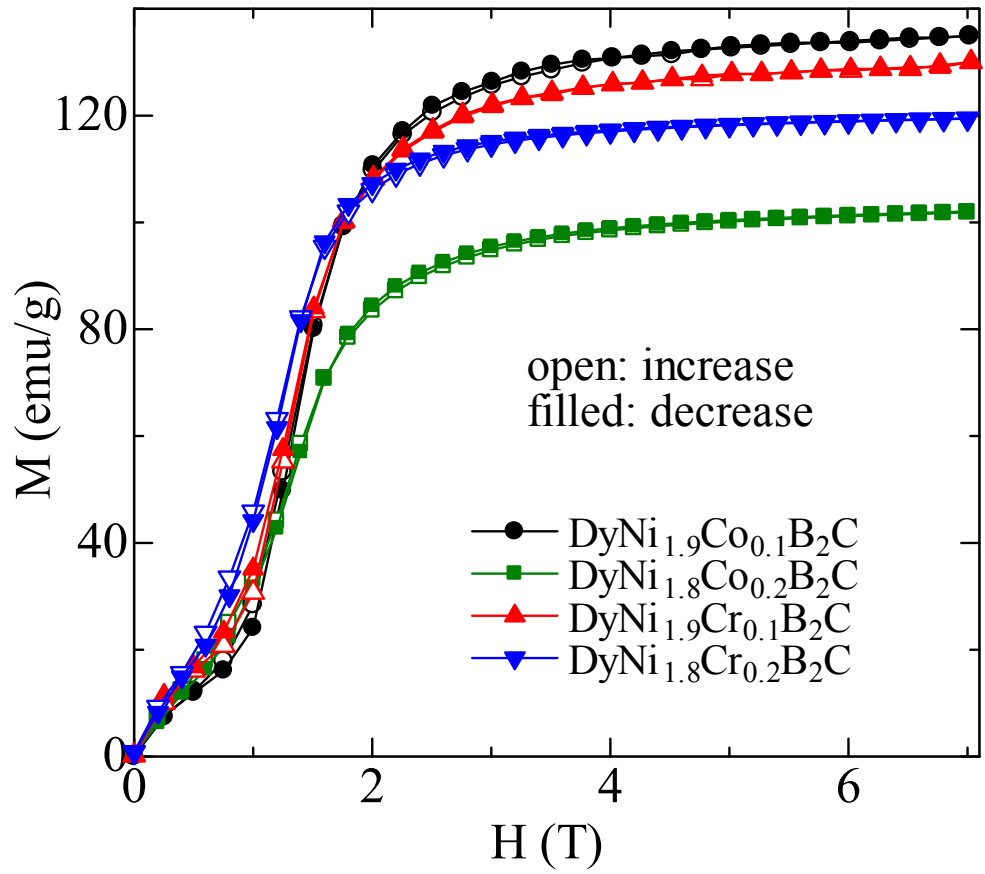

Fig. 21.

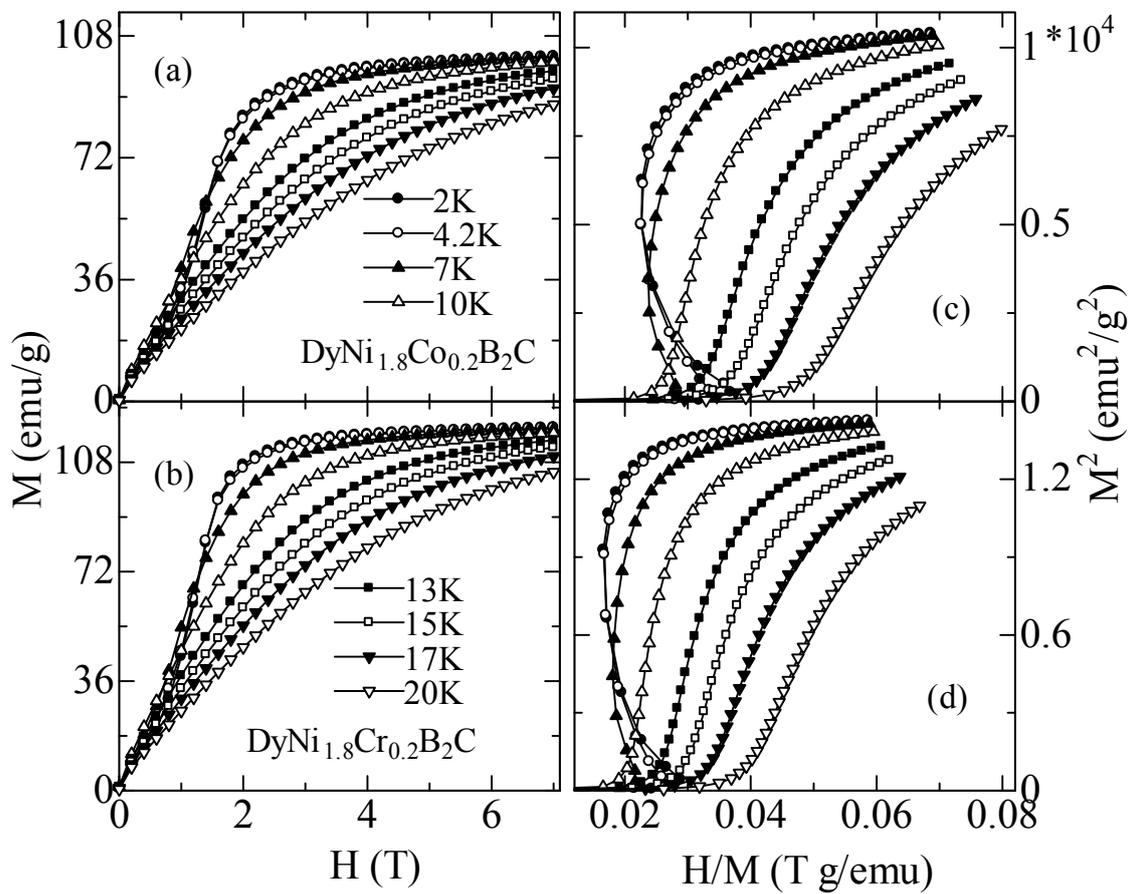

Fig. 22.

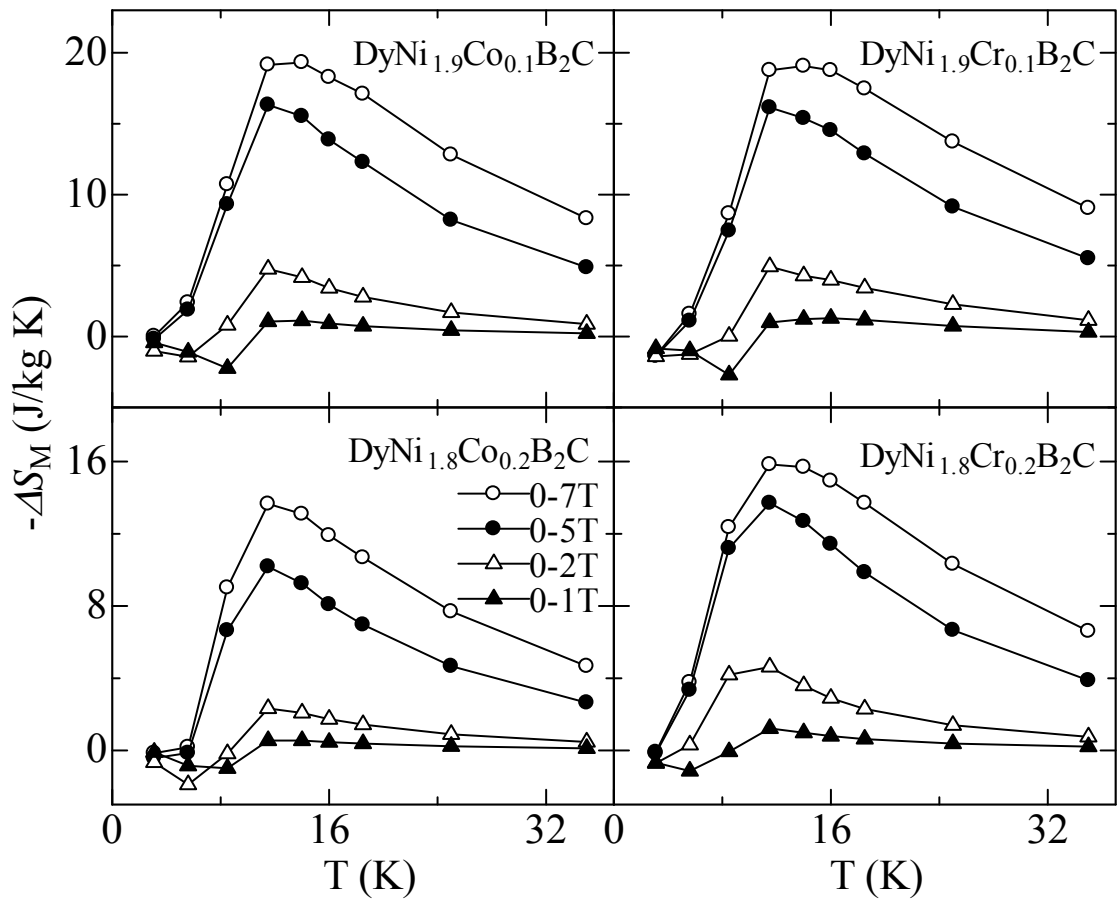

Fig. 23.

Table I. The transition temperature ($T_M$), the maximum magnetic entropy change ($-\Delta S_M^{max}$), maximum adiabatic temperature change ($\Delta T_{ad}^{max}$), and relative cooling power (*RCP*) under the field changes of 0-2 T and 0-5 T for the intermetallic compounds of rare earth with low boiling point metal(s) (Zn, Mg, and Cd) as well as some giant/large MCE materials with similar $T_M$. The column with a mark "--" means that the value was not reported in the literature.

| Compound | $T_M$ (K) | $-\Delta S_M^{max}$ (J /kg K) | | RCP (J/kg) | | $\Delta T_{ad}^{max}$ (K) | | Ref. |
|---|---|---|---|---|---|---|---|---|
| | | 2 T | 5 T | 2 T | 5 T | 2T | 5T | |
| TmZn | 8.4 | 19.6 | 26.9 | 76 | 269 | 3.3 | 8.6 | 57 |
| HoZn | 26/72 | 6.5 | 12.1 | 255 | 792 | -- | -- | 58 |
| EuAuZn | 51 | 4.8 | 9.1 | 105 | 318 | -- | -- | 59 |
| EuRh$_{1.2}$Zn$_{0.8}$ | 95 | 3.1 | 5.7 | 233 | 513 | 1.4 | 2.8 | 60 |
| TmZnAl | 2.8 | 4.3 | 9.4 | 53 | 189 | -- | -- | 61 |
| Eu$_4$PdMg | 150 | 2.6 | 5.5 | 988 | 1346 | -- | -- | 66 |
| Er$_4$PtMg | 15 | 8.5 | 17.9 | 152 | 483 | 1.8 | 4.1 | 67 |
| Er$_4$PdMg | 21 | 6.2 | 15.5 | 142 | 457 | 1.7 | 3.7 | 68 |
| Ho$_4$PtMg | 28 | 6.1 | 13.4 | 177 | 527 | 2.2 | 5.0 | 67 |

| Compound | | | | | | | | |
|---|---|---|---|---|---|---|---|---|
| Gd$_2$Ni$_{0.5}$Cu$_{1.5}$Mg | 57 | 5.3 | 9.5 | -- | 688 | 1.6 | 3.2 | 69 |
| Gd$_2$NiCuMg | 55.2 | 6.1 | 11.4 | 202 | 630 | -- | -- | 69 |
| Gd$_4$Co$_2$Mg$_3$ | 77 | 5.8 | -- | -- | -- | 1.3 | -- | 70 |
| Er$_4$NiCd | 5.9 | 7.3 | 18.3 | 237 | 595 | 3.1 | 7.7 | 72 |
| GdCd$_{0.9}$Ru$_{0.1}$ | 149 | 1.8 | 4.1 | 235 | 636 | -- | -- | 73 |
| GdCd$_{0.85}$Ru$_{0.15}$ | 108 | 2.4 | 5.8 | 223 | 597 | -- | -- | 73 |
| GdCd$_{0.8}$Ru$_{0.2}$ | 73 | 3.8 | 8.5 | 213 | 583 | -- | -- | 73 |
| ErCr$_2$Si$_2$ | 1.9 | 24.1 | 29.7 | 136 | 388 | 8.4 | 17.4 | 76 |
| TmAgAl | 3.3 | 7.1 | 12.4 | 74 | 214 | -- | -- | 61 |
| GdCr$_2$Si$_2$ | 4.5 | 3.1 | 14.1 | 35 | 212 | -- | -- | 77 |
| ErNiBC | 5 | 17.1 | 24.8 | 155 | 312 | 5.3 | 8.6 | 78,79 |
| TmMn$_2$Si$_2$ | 5.5 | 15.3 | 22.7 | 106 | 250 | 5.0 | 10.1 | 80 |
| ErMn$_2$Si$_2$ | 6.5 | 20.0 | 25.2 | 130 | 365 | 5.4 | 12.9 | 81 |
| Tm$_3$Co | 8 | 11.6 | 19.9 | 93 | 300 | -- | -- | 82 |
| HoCo$_2$B$_2$ | 10 | 6.8 | 12.2 | 83 | 271 | -- | -- | 83 |
| DyCo$_2$B$_2$ | 10 | 5.4 | 12.1 | 88 | 282 | -- | -- | 83 |
| ErAgAl | 14 | 4.2 | 10.5 | 77 | 261 | -- | -- | 84 |
| TbCo$_2$B$_2$ | 15 | 1.5 | 6.2 | 49 | 292 | -- | -- | 83 |
| GdNiBC | 15 | 9.3 | 19.8 | 188 | 474 | 4.2 | 9.9 | 78,85 |
| HoAgAl | 18 | 3.8 | 10.3 | 99 | 344 | -- | -- | 84 |
| DyCo$_3$B$_2$ | 22 | 7.4 | 12.6 | 154 | 397 | 6.4 | 11.6 | 86 |
| HoPdIn | 6/23 | 7.9 | 14.6 | 112 | 496 | 2.6 | 5.5 | 87 |
| GdCo$_2$B$_2$ | 25 | 9.3 | 17.1 | 167 | 462 | -- | -- | 88 |
| TbCo$_3$B$_2$ | 28 | 4.9 | 8.7 | 59 | 295 | 4.0 | 7.3 | 89 |
| NdMn$_2$Ge$_{0.4}$Si$_{1.6}$ | 36 | 12.3 | 28.4 | 95 | 284 | 2.1 | 4.1 | 90 |
| GdCo$_3$B$_2$ | 54 | 5.0 | 9.4 | 110 | 357 | 2.8 | 5.2 | 91 |
| ErFeAl | 55 | 2.4 | 6.1 | 77 | 311 | -- | -- | 92 |
| DyFeSi | 70 | 9.2 | 17.4 | -- | -- | 3.4 | 7.1 | 93 |
| HoFeAl | 80 | 3.4 | 7.5 | 174 | 563 | -- | -- | 92 |
| Ho$_2$In | 85 | -- | 11.2 | -- | 360 | -- | -- | 94 |
| Gd$_{53}$Al$_{24}$Co$_{20}$Zr$_3$ (micro-wire) | 96 | -- | 10.3 | -- | 733 | -- | -- | 95 |
| Gd$_{55}$Al$_{20}$Co$_{20}$Ni$_5$ | 105 | 6.59 | 9.8 | -- | 829 | -- | 4.74 | 96 |

| Material | (metallic glass) | | | | | | | |
|---|---|---|---|---|---|---|---|---|
| DyFeAl | 126 | 3.1 | 6.4 | 190 | 595 | -- | -- | 97 |
| Gd$_5$Ir$_2$Sn | 154 | 3.9 | 7.3 | 176 | 423 | -- | -- | 98 |

Table II. The superconductivity transition temperature ($T_{SC}$), magnetic transition temperature ($T_M$), the maximum values of magnetocaloric parameters (-$\Delta S_M^{max}$ and $\Delta T_{ad}$) as well as the relative cooling power (RCP) for the magnetic field change of 0-5 T in the rare earth nickel boroncarbides $RE$Ni$_2$B$_2$C and $RE$NiBC compounds.

The column with a mark "--" means that the value was not reported in the literature.

The column with a mark "N" means that superconductivity was not observed above 2 K.

| Material | $T_{SC}$ (K) | $T_M$ (K) | -$\Delta S_M^{max}$ (J/kg K) | $\Delta T_{ad}^{max}$ (K) | RCP (J/kg) | Reference |
|---|---|---|---|---|---|---|
| Dy$_{0.9}$Tm$_{0.1}$Ni$_2$B$_2$C | 4.5 | 9.2 | 14.7 | -- | 248 | 104 |
| HoNi$_2$B$_2$C | 8.2 | 5 | 17.7/19.2 | 11 | 283 | 105,106 |
| Dy$_{0.3}$Ho$_{0.7}$Ni$_2$B$_2$C | 8.1 | 6 | 18.7 | -- | 264 | 106 |
| Dy$_{0.5}$Ho$_{0.5}$Ni$_2$B$_2$C | 6.2 | 8 | 18.5 | -- | 275 | 106 |

| Compound | | | | | | |
|---|---|---|---|---|---|---|
| Dy$_{0.7}$Ho$_{0.3}$Ni$_2$B$_2$C | 6.4 | 8.5 | 20.2 | -- | 243 | 106 |
| DyNi$_2$B$_2$C | 6.4 | 10.5 | 17.6/17.1 | 9.7 | 290 | 105,106 |
| DyNi$_{1.9}$Co$_{0.1}$B$_2$C | N | 8.4 | 16.3 | -- | 309 | 107 |
| DyNi$_{1.9}$Co$_{0.2}$B$_2$C | N | 8 | 10.2 | -- | 168 | 107 |
| DyNi$_{1.9}$Cr$_{0.1}$B$_2$C | N | 9.2 | 16.1 | -- | 272 | 107 |
| DyNi$_{1.9}$Cr$_{0.2}$B$_2$C | N | 8.8 | 13.7 | -- | 219 | 107 |
| ErNi$_2$B$_2$C | 10.5 | 6.1 | 11.1/9.8 | 4.6 | 155 | 105,108 |
| ErNi$_{1.9}$Fe$_{0.1}$B$_2$C | N | 5.6 | 9.6 | -- | 106 | 108 |
| ErNi$_{1.9}$Fe$_{0.2}$B$_2$C | N | 5.2 | 8.0 | -- | 71 | 108 |